\documentclass[a4paper,11pt]{article}
\usepackage{jheppub} 

\usepackage[utf8]{inputenc}
\usepackage[colorlinks=true,citecolor=blue,linkcolor=blue]{hyperref}
\usepackage[normalem]{ulem}
\usepackage{amsmath,amssymb}
\usepackage{epsfig}
\usepackage{graphicx}  % Standard graphics package 
\usepackage{placeins} 
\usepackage{url}
\usepackage{color}
\usepackage{slashed}
\usepackage{multirow}
\usepackage{placeins}
\usepackage[dvipsnames]{xcolor}
\usepackage{epstopdf}
\usepackage{soul}
\usepackage{tikz}
\usepackage[capitalise, english]{cleveref}
\usepackage{siunitx}
\usepackage{xspace}
\usetikzlibrary{trees}
\usepackage{comment}
\usetikzlibrary{decorations.pathmorphing}
\usetikzlibrary{decorations.markings}
\usetikzlibrary{calc,tikzmark,fit,shapes.geometric,matrix,decorations.markings,arrows.meta,decorations.pathmorphing,patterns,positioning,snakes}

\tikzset
  {midarrow/.style={decoration={markings,mark=at position 0.5 with
     {\arrow[thin,xshift=2pt]{Triangle[length=4pt,#1]}}},postaction={decorate}}
  }

\tikzset{
proton/.style = {circle, draw=black, thin, fill=black!20!white, minimum size=#1,
              inner sep=0pt, outer sep=0pt},
proton/.default = 6pt % size of the circle diameter 
}

\tikzset{
blob/.style = {circle, draw=black, thin, preaction={fill, black!20!white}, pattern=north east lines, minimum size=#1,
              inner sep=0pt, outer sep=0pt},
blob/.default = 6pt % size of the circle diameter 
}

\tikzset{
wc/.style = {circle, fill, minimum size=#1,
              inner sep=0pt, outer sep=0pt},
wc/.default = 4pt % size of the circle diameter 
}

\tikzset{vector/.style={decorate, decoration=snake}}

\newcommand\myshade{80}
\colorlet{mylinkcolor}{ForestGreen}
\colorlet{mycitecolor}{Red}
\colorlet{myurlcolor}{violet}

\hypersetup{
  linkcolor  = mylinkcolor!\myshade!black,
  citecolor  = mycitecolor!\myshade!black,
  urlcolor   = myurlcolor!\myshade!black,
  colorlinks = true
}
\definecolor{jblue}{RGB}{20,50,100}
\definecolor{npurple}{RGB} {153, 51, 204}
\definecolor{wred}{RGB}{217,0,56}
\definecolor{white}{RGB}{255,255,255}

\definecolor{korange}{RGB}{235, 80,  43}
\definecolor{korange2}{RGB}{245, 100,  63}
\definecolor{kyelloworange}{RGB}{255, 210,  110}
\definecolor{kyelloworange2}{RGB}{240, 170,  90}
\definecolor{kred}{RGB}{204,  102, 153}
\definecolor{kpurple}{RGB}{153,  61, 190}
\definecolor{kpurplelight}{RGB}{213,  161, 230}

\allowdisplaybreaks

\newcommand{\cC}{\mathcal{C}}

\newcommand{\cB}{\mathcal{B}}
\newcommand{\cL}{\mathcal{L}}
\newcommand{\cO}{Q}
\newcommand{\cOp}{\mathcal{O}}
\newcommand{\epsf}{\varepsilon_{{}_{F}}}
\newcommand{\epsq}{\varepsilon_{{}_{Q}}}
\newcommand{\epsl}{\varepsilon_{{}_{L}}}
\newcommand{\epsh}{\varepsilon_{{}_{H}}}
\newcommand{\epsloop}{\varepsilon_{{}_{\rm loop}}}

\makeatletter 
\gdef\@fpheader{}
\makeatother
\begin{document}

%=============================================================================

\title{New Physics in the Third Generation:
A Comprehensive SMEFT Analysis and Future Prospects }

\author[a]{Lukas Allwicher,}
\emailAdd{lukas.allwicher@physik.uzh.ch}

\author[b]{Claudia Cornella,}
\emailAdd{claudia.cornella@uni-mainz.de}

\author[a]{Gino Isidori,}
\emailAdd{isidori@physik.uzh.ch}

\author[c]{and Ben A. Stefanek}
\emailAdd{benjamin.stefanek@kcl.ac.uk}

\affiliation[a]{Physik-Institut, Universit\"at Z\"urich, CH-8057 Z\"urich, Switzerland}

\affiliation[b]{PRISMA$^+$ Cluster of Excellence {\em \&} MITP, 
Johannes Gutenberg University, Mainz, Germany}

\affiliation[c]{Physics Department, King’s College London, Strand, London, WC2R 2LS, United Kingdom}

\date{\today}

\preprint{ZU-TH 71/23, MITP-23-060, KCL-PH-TH/2023-59}

%=============================================================================

\abstract{We present a comprehensive analysis of electroweak, 
flavor, and collider bounds on the complete set of 
dimension-six SMEFT operators in the $U(2)^5$-symmetric limit.
This operator basis provides a consistent framework to describe 
 a wide class of new physics models and, in particular, the motivated class of models  where the new degrees of freedom couple mostly to the third generation. 
By analyzing observables from all three sectors, and consistently including renormalization group evolution, we provide bounds on the effective scale of all 124 $U(2)^5$-invariant operators.
The relation between flavor-conserving and flavor-violating observables 
is analyzed taking into account the leading $U(2)^5$ 
breaking in the Yukawa sector, which is responsible for heavy-light quark mixing. 
We show that under simple, motivated, and non-tuned hypotheses 
for the parametric size of the Wilson coefficients at the high scale, 
 {\em all} present bounds are consistent with an effective scale 
 as low as 1.5~TeV. We also show that a future circular $e^+ e^-$ collider program such as FCC-ee would 
push most of these bounds by an order of magnitude. 
This would rule out or provide clear evidence
for a wide class of  compelling 
new physics models that are fully compatible with present data. 
}

\maketitle

%-----------------------------------------------------------------------------
\section{Introduction}

Several arguments, ranging from cosmological observations to the instability of the Higgs mass, indicate that the Standard Model (SM) has a  
non-trivial ultraviolet (UV) completion. Above a certain energy threshold, we expect new dynamical degrees of freedom, generically denoted as new physics (NP),  to emerge.  Pinpointing  this energy threshold is arguably the most pressing and debated question in particle physics nowadays. 

A natural framework to systematically address this question is provided by the so-called Standard Model Effective Field Theory, or SMEFT
(see~\cite{Brivio:2017vri,Isidori:2023pyp,Falkowski:2023hsg} for recent reviews). Within the SMEFT, the effects of new heavy dynamics are encoded in a tower of higher-dimensional operators constructed using SM fields. By comparing data to theory across various sectors (e.g.~electroweak, flavor, and collider observables), bounds can be set on the Wilson coefficients (WCs) of these operators. These can in turn be interpreted as lower bounds on the effective cut-off scale of the theory, in the limit of $O(1)$ couplings.  In the absence of additional dynamical assumptions, these bounds span several orders of magnitude: leaving aside the special case of baryon- and lepton-number violating operators, the bounds range from  $O(100)$~GeV to $O(10^5)$~TeV.

However, interpreting these constraints as strict lower limits for the energy thresholds of new dynamics can be misleading. In many motivated extensions of the SM there are several operators, especially the flavor-violating ones, that have suppressed couplings and thus play only a marginal role.
Indeed, this is exemplified a  Yukawa sector of the SM itself, but it is not captured within the SMEFT without further assumptions.

A more informative approach to gain insight into motivated classes of SM extensions 
is to proceed via the following three-step strategy~\cite{Isidori:2023pyp}: 
i)~identify the main features of a given class of NP models (typically expressed in terms of exact or approximate symmetries) and translate them into a series of   selection rules on SMEFT operators in the UV, 
ii)~take into account the renormalization group evolution (RGE) down to the electroweak scale, iii)~evaluate observables and compare with experimental data.
We emphasize that only a consistent implementation of these three different steps
allows for a meaningful comparison of the bounds derived from electroweak, flavor, and collider data. 

Following this approach, in this paper we present a general analysis of SM extensions in which NP couples universally to 
the two lightest fermion families, while it might couple differently to the third one.  
Specifically, we analyze models where the corresponding dimension-six SMEFT operators are invariant under a $U(2)^5$ flavor symmetry~\cite{Barbieri:2011ci,Barbieri:2012uh,Isidori:2012ts} acting on the lightest two SM families. Namely, we impose a UV partitioning of the SMEFT, separating operators acting on the third and light generations, and requiring that the non-degeneracy among the light families is confined to the small $U(2)^5$-breaking terms already  present in the SM Yukawa couplings. 

A special role for the third family is a feature common to several  well-motivated  NP models. This is closely related to the fact that the top-quark Yukawa coupling is the largest couplings of the Higgs field
(and of the whole SM) at the electroweak (EW) scale. 
At the same time, the requirement of quasi-degenerate NP couplings to the light families follows from the non-observation of significant deviations from SM predictions in  flavor-changing processes.
Not surprisingly, the $U(2)^5$ flavor symmetry has been discussed both in the context of supersymmetric models 
(see e.g.~\cite{Barbieri:2011ci,Papucci:2011wy,Larsen:2012rq}), 
composite models 
(see e.g.~\cite{Barbieri:2012tu,Matsedonskyi:2014iha,Panico:2016ull}) 
and, more recently, in the framework  of flavor non-universal gauge models
(see e.g.~\cite{Bordone:2017bld,Greljo:2018tuh,Fuentes-Martin:2022xnb,Davighi:2023iks,FernandezNavarro:2023rhv,Davighi:2023evx}). 
The effective theory we are considering is therefore representative of a wide class of  compelling  NP models.

Our analysis serves two primary objectives.  First, we seek to understand how low the NP scale can be under the additional dynamical hypothesis of NP coupling dominantly to third-generation fermions.  It is important to emphasize that the $U(2)^5$ flavor symmetry, by itself, does not dictate whether NP couples mainly to light or heavy fermions. Instead, it only provides a consistent way to differentiate these two sectors in the UV. Dominant NP couplings to the third generation can be achieved by imposing additional conditions on the parametric size of the WCs at the high scale. Our goal is to determine these conditions trying to be as general as possible, examining the bounds on all 124 CP-even independent $U(2)^5$-invariant operators present in the SMEFT at $d=6$~\cite{Faroughy:2020ina}. We analyze the bounds on each operator separately,  thus avoiding flat directions related to cancellations which might appear in specific models. This approach ensures a conservative estimation of the NP scale as well as the identification of general high-scale dynamical conditions that permit this scale to be low.
  Our findings reveal that, under a few simple and rather natural hypotheses, the bounds on all operators are consistent with an effective scale as low as 1.5~TeV, which aligns with the value deduced by collider constraints on third-generation contact interactions. This result is very encouraging in view of on-going experimental efforts, both at the low- and the high-energy frontier. 

Our second key objective is to explore potential improvements of these bounds in the future, especially considering the prospect of a future $e^+ e^-$ collider such as the FCC-ee program.   Although last revised over 20 years ago at LEP, many of these bounds are still dominated by electroweak precision observables (EWPO). Given this, the FCC-ee program is expected to have an enormous impact.
 In this context it is  important to emphasize that while the $U(2)^5$ flavor symmetry   effectively sidesteps  flavor and collider bounds, it is less effective  against EWPO. This is particularly true if we combine the
$U(2)^5$ symmetry with the hypothesis of NP coupled mainly to the third generation,  exploiting the inherent asymmetry between heavy and light flavors at proton-proton colliders.  This asymmetry is not present in many EWPO  given their sensitivity to heavy flavors, either at the tree level or via  quantum effects.
Given these considerations, a future EW precision machine like FCC-ee stands out as the optimal method to either confirm or refute a range of compelling NP models that are currently consistent with present data.

Important progress is also expected on a shorter timescale
in view of the improvements foreseen  for selected flavor-violating observables. In our framework, all flavor-violating effects originate   from  the off-diagonal entries of the SM Yukawa couplings. However, unlike in the Minimal Flavor Violation (MFV) hypothesis~\cite{DAmbrosio:2002vsn}, there is an additional breaking of flavor universality related to the special role attributed to the third generation. The phenomenological consequences of this  additional breaking of universality  depend on the orientation, in flavor space, of what we denote as third generation in the underlying model (i.e.~in the higher-dimensional operators). 
To probe this effect in sufficient generality we  allow  the orientation of the third-generation quark doublet 
to vary between the down-alignment limit, where it is identified with the $b$-quark $SU(2)_L$ doublet, and an up-alignment limit, where it is identified with the top-quark $SU(2)_L$ doublet. 

The paper is organized as follows. In Section~\ref{sect:basis}
we define the basis of the $d=6$ operators, with emphasis on their flavor structure. In Section~\ref{sect:bounds} we present the bounds on the WCs obtained from EW precision, flavor, and collider data, without additional dynamical hypotheses. In Section~\ref{sect:third} we focus on NP coupled dominantly to the third generation, defining the additional UV conditions related to this hypothesis, followed by a comparison to the data.
 Section~\ref{sect:future} is devoted to future projections: we first illustrate a specific example of the interplay between flavor, collider, and EWPO, and then we analyze the impact of a future circular $e^+ e^-$ collider program such as FCC-ee on the most relevant EWPO. 
Finally, our results are summarized in the Conclusions.

%----------------------------------------------------------------------------
\section{The operator basis}
\label{sect:basis}
%----------------------------------------------------------------------------

The first complete, non-redundant basis for the SMEFT at $d=6$
has been presented in~\cite{Grzadkowski:2010es}: it consists of 
59 independent gauge-invariant structures, which reduce 
to 53 if we exclude  CP-odd terms.  This relatively small number of terms gives rise to a large proliferation of independent operators (1350 in the CP-conserving limit) 
once flavor indices are incorporated. Flavor symmetries therefore play a key role in both restricting and organizing this otherwise large basis~\cite{Faroughy:2020ina,Greljo:2022cah}. As anticipated, in this work 
we focus on $U(2)^5$-invariant operators.   

 The $U(2)^5 = U(2)_q \times U(2)_u \times U(2)_d \times U(2)_\ell \times U(2)_e$ symmetry is a global symmetry acting on each of the five SM fermion species with independent gauge quantum numbers ($q,u,d,\ell, e$). 
Within  each of these species, the two lightest generations transform as a doublet under a specific subgroup of $U(2)^5$, whereas the third generation is a singlet.    
The $d=6$  SMEFT operators in the exact $U(2)^5$ symmetric limit have been classified in~\cite{Faroughy:2020ina}. In the CP-conserving limit, they consist of 124 independent terms.  A concise representation of these terms can be found in  Table~\ref{tab:Warsaw-basis}, and is subsequently made more explicit in Tables~\ref{tab:psi2H2D}--\ref{tab:bosonic}. 
The effective Lagrangian we are considering can thus be decomposed as 
\begin{equation}
\mathcal{L} = \cL_{\rm SM}^{(d=4)}+ \sum_i \cC_{i} \cO_{i} \,,
\end{equation}
where the $\cO_i$ are the operators listed in 
Tables~\ref{tab:psi2H2D}--\ref{tab:bosonic}.  

The following points allow to easily understand how the full SMEFT basis is reduced to the $U(2)^5$-invariant 
subset, and clarify our notation for flavor indices:
\begin{itemize} 
\item[i] Bosonic operators are not affected by flavor symmetries.
\item[ii.] Non-Hermitian operators necessarily involve different fermion species, hence in the exact $U(2)^5$ limit they are allowed only for third-generation fermions.
\item[iii.] Hermitian fermion bilinears always appear in two forms: with two third-generation fermions, and with two light-generation fermions.  For instance, the operator $Q_{He}$ gives rise to the following two  
 $U(2)^5$-invariant  operators:
\begin{equation}
  Q_{He}^{[33]}=(H^\dagger iD_\mu H)(\bar e_3 \gamma^\mu e_3)\,, \qquad 
  Q_{He}^{[ii]}=  (H^\dagger iD_\mu H) \sum_{i=1,2} (\bar e_i \gamma^\mu e_i)\,.
\end{equation}
\item[iv.] Hermitian four-fermion operators involving two different fermion species appear in four distinct forms. As an example,  the operator 
$Q_{\ell q}$ gives rise to:
\begin{align}
&Q_{\ell q}^{(1)[3333]} =  (\bar\ell_3 \gamma_\mu \ell_3)(\bar q_3 \gamma^\mu q_3)\,, \quad 
&Q_{\ell q}^{(1)[iijj]} =  \sum_{i=1,2} (   \bar\ell_i \gamma_\mu \ell_i ) \sum_{j=1,2}  ( \bar q_j \gamma^\mu q_j )\,, 
\nonumber \\
&Q_{\ell q}^{(1)[33ii]} =  (\bar\ell_3 \gamma_\mu \ell_3)  \sum_{i=1,2} (  \bar q_i \gamma^\mu q_i )\,, \quad 
&Q_{\ell q}^{(1)[ii33]} =  \sum_{i=1,2} ( \bar\ell_i \gamma_\mu \ell_i)(\bar q_3 \gamma^\mu q_3)\,. \qquad 
 \end{align}
\item[v.] Hermitian four-fermion operators involving a single fermion species (hence four identical fields but for the flavor index) can appear in up to five different   configurations.
\end{itemize}

\begin{table}[p]
\small{
	\newcommand{\OpScale}{.85} 
	\renewcommand{\arraystretch}{1.5}
    \centering 
	\scalebox{\OpScale}{%%%%%
	\setlength{\tabcolsep}{0.8mm}
    \hspace{1.3cm}
	\begin{tabular}{| rl | rl | }
		\multicolumn{4}{c}{ Hermitian $\psi^2$ operators:~
        $7_{[\rm GIS]} \to 14_{[U(2)^5]}$~  (Table~\ref{tab:psi2H2D})  } \\[.1cm] \hline 
		\,
		$Q_{H\ell}^{(1)[pr]}=$ & $(H^\dagger i \overleftrightarrow{D}_\mu H)(\bar\ell_p \gamma^\mu \ell_r)$ & \,
		$Q_{H e}^{[pr]}=$ & $(H^\dagger i \overleftrightarrow{D}_\mu H)(\bar e_p \gamma^\mu e_r)$  \\
		$Q_{H\ell}^{(3)[pr]}=$ & $(H^\dagger i \overleftrightarrow{D}_\mu^I H)(\bar\ell_p \tau^I\gamma^\mu \ell_r)$ \quad & 
		$Q_{H u}^{[pr]}=$ & $(H^\dagger i \overleftrightarrow{D}_\mu H)(\bar u_p \gamma^\mu u_r)$  \\
		$Q_{Hq}^{(1)[pr]}=$ & $(H^\dagger i \overleftrightarrow{D}_\mu H)(\bar q_p \gamma^\mu q_r)$ & 
		$Q_{H d}^{[pr]}=$ & $(H^\dagger i \overleftrightarrow{D}_\mu H)(\bar d_p \gamma^\mu d_r)$ \quad \\
		$Q_{Hq}^{(3)[pr]}=$ & $(H^\dagger i \overleftrightarrow{D}_\mu^I H)(\bar q_p \tau^I\gamma^\mu q_r)$  & &
        \\ \hline
	\end{tabular}
	}%%%%%
	\vspace{0.3cm} \newline \centering 
	\scalebox{\OpScale}{%%%%%
 	\setlength{\tabcolsep}{0.8mm}
	\centering
	\begin{tabular}{| rl || rl | rl | rl |}
		\multicolumn{8}{c}{ Non-hermitian $\psi^2$ operators:~
        $12_{[\rm GIS]} \to 12_{[U(2)^5]}$~ 
        (Table~\ref{tab:dipole})  } \\[.1cm] \hline
        \,
		$Q_{eH}^{[pr]}=$ & $(H^\dagger H)(\bar\ell_p e_r H)$ \, &	\,
		$Q_{uG}^{[pr]}=$ & $(\bar q_p \sigma^{\mu\nu}T^A u_r)\widetilde{H}G_{\mu\nu}^A$ \, &  \, 
		$Q_{dG}^{[pr]}=$ & $(\bar q_p \sigma^{\mu\nu}T^A d_r)H G_{\mu\nu}^A$\,  & 
        $Q_{eW}^{[pr]}=$ & $(\bar\ell_p \sigma^{\mu\nu}e_r)\tau^I H W_{\mu\nu}^I$ \\ 
		$Q_{uH}^{[pr]}=$ & $(H^\dagger H)(\bar q_p u_r \widetilde{H})$ &
		$Q_{uW}^{[pr]}=$ & $(\bar q_p \sigma^{\mu\nu}u_r)\tau^I \widetilde{H}W_{\mu\nu}^I$ &
		$Q_{dW}^{[pr]}=$ & $(\bar q_p \sigma^{\mu\nu}d_r)\tau^I H W_{\mu\nu}^I$ &
        $Q_{eB}^{[pr]}=$ & $(\bar\ell_p \sigma^{\mu\nu}e_r) H B_{\mu\nu}$  \\  \cline{7-8} 
		$Q_{dH}^{[pr]}=$ & $(H^\dagger H)(\bar q_p d_r H)$ &
		$Q_{uB}^{[pr]}=$ & $(\bar q_p \sigma^{\mu\nu}u_r)\widetilde{H}B_{\mu\nu}$ &
		$Q_{dB}^{[pr]}=$ & $(\bar q_p \sigma^{\mu\nu}d_r) H B_{\mu\nu}$  & \,
		$Q_{Hud}^{[pr]}=$ & $i(\widetilde{H}^\dagger D_\mu H)(\bar u_p \gamma^\mu d_r)$ \\ \hline
		\end{tabular}
	}%%%%%
	\vspace{0.3cm}
	\\ \centering
	\scalebox{\OpScale}{%%%%%
	\setlength{\tabcolsep}{0.8mm}
	\begin{tabular}{| rl | rl | rl |}
		\multicolumn{6}{c}{ Hermitian $\psi^4$ operators:~ 
        $20_{[\rm GIS]} \to 84_{[U(2)^5]}$~
        (Tables~\ref{tab:LLLL}--\ref{tab:LLRR}) } \\[.1cm] \hline
		\multicolumn{2}{|c|}{$(\bar L L)(\bar L L)$:\quad  $5_{[\rm GIS]} \to 23_{[U(2)^5]}$ } & 
		\multicolumn{2}{c|}{$(\bar R R)(\bar R R)$:\quad  $7_{[\rm GIS]} \to 29_{[U(2)^5]}$ }  & 
		\multicolumn{2}{c|}{$(\bar L L)(\bar R R)$:\quad   $8_{[\rm GIS]} \to 32_{[U(2)^5]}$ } 
		\\ \hline
		$Q_{\ell\ell}^{[prst]}=$ & $(\bar\ell_p \gamma_\mu \ell_r)(\bar\ell_s \gamma^\mu \ell_t)$ & 
		$Q_{ee}^{[prst]}=$ & $(\bar e_p \gamma_\mu e_r)(\bar e_s \gamma^\mu e_t)$ & 
		$Q_{\ell e}^{[prst]}=$ & $(\bar\ell_p \gamma_\mu \ell_r)(\bar e_s \gamma^\mu e_t)$  \\
		$Q_{qq}^{(1)[prst]}=$ & $(\bar q_p \gamma_\mu q_r)(\bar q_s \gamma^\mu q_t)$ & 
		$Q_{uu}^{[prst]}=$ & $(\bar u_p \gamma_\mu u_r)(\bar u_s \gamma^\mu u_t)$ & 
		$Q_{\ell u}^{[prst]}=$ & $(\bar\ell_p \gamma_\mu \ell_r)(\bar u_s \gamma^\mu u_t)$  \\
         \,
		$Q_{qq}^{(3)[prst]}=$ & $(\bar q_p \gamma_\mu \tau^I q_r)(\bar q_s \gamma^\mu \tau^I q_t)$ \, & 
		$Q_{dd}^{[prst]}=$ & $(\bar d_p \gamma_\mu d_r)(\bar d_s \gamma^\mu d_t)$ & 
		$Q_{\ell d}^{[prst]}=$ & $(\bar\ell_p \gamma_\mu \ell_r)(\bar d_s \gamma^\mu d_t)$  \\
		$Q_{\ell q}^{(1)[prst]}=$ & $(\bar\ell_p \gamma_\mu \ell_r)(\bar q_s \gamma^\mu q_t)$ & 
		$Q_{eu}^{[prst]}=$ & $(\bar e_p \gamma_\mu e_r)(\bar u_s \gamma^\mu u_t)$ & 
		$Q_{q e}^{[prst]}=$ & $(\bar q_p \gamma_\mu q_r)(\bar e_s \gamma^\mu e_t)$\\
		$Q_{\ell q}^{(3)[prst]}=$ & $(\bar\ell_p \gamma_\mu \tau^I \ell_r)(\bar q_s \gamma^\mu \tau^I q_t)$ & 
		$Q_{ed}^{[prst]}=$ & $(\bar e_p \gamma_\mu e_r)(\bar d_s \gamma^\mu d_t)$ & 
		$Q_{qu}^{(1)[prst]}=$ & $(\bar q_p \gamma_\mu q_r)(\bar u_s \gamma^\mu u_t)$ \\ 
		& & 
		$Q_{ud}^{(1)[prst]}=$ & $(\bar u_p \gamma_\mu u_r)(\bar d_s \gamma^\mu d_t)$ & \,
		$Q_{qu}^{(8)[prst]}=$ & $(\bar q_p \gamma_\mu T^A q_r)(\bar u_s \gamma^\mu T^A u_t)$ \, \\ 
		& & 
        \, 
		$Q_{ud}^{(8)[prst]}=$ & $(\bar u_p \gamma_\mu T^A u_r)(\bar d_s \gamma^\mu T^A d_t)$ \, & 
		$Q_{qd}^{(1)[prst]}=$ & $(\bar q_p \gamma_\mu q_r)(\bar d_s \gamma^\mu d_t)$   \\  
		& & 
		& &   \,
		$Q_{qd}^{(8)[prst]}=$ & $(\bar q_p \gamma_\mu T^A q_r)(\bar d_s \gamma^\mu T^A d_t)$ \, \\ \hline
	\end{tabular}
 	}
	\vspace{0.3cm}
	\\ \centering
	\scalebox{\OpScale}{%%%%%
	\setlength{\tabcolsep}{0.8mm}
	\begin{tabular}{| r l | r l | r l |}
		\multicolumn{6}{c}{ Non-hermitian $\psi^4$ operators:~ 
        $5_{[\rm GIS]} \to 5_{[U(2)^5]}$~ (Table~\ref{tab:nhpsi4}) } \\[.1cm] \hline  \,
    $Q_{\ell equ}^{(1)[prst]}=$ & $(\bar \ell_p^i e_r)\varepsilon_{ij}(\bar q_s^j u_t)$ &  $
    Q_{quqd}^{(1)[prst]}=$ & $(\bar q_p^i u_r)\varepsilon_{ij}(\bar q_s^j d_t)$ & \,
	$Q_{\ell e d q }^{[prst]}=$ & $(\bar \ell _p^i e_r)(\bar d_s q_{ti})$ \,  \\  \,
	$Q_{\ell equ}^{(3)[prst]}=$ & $(\bar \ell_p^i \sigma_{\mu\nu} e_r)\varepsilon_{ij}(\bar q_s^j \sigma^{\mu\nu} u_t)$  
    &	\, $Q_{quqd}^{(8)[prst]}=$ & $(\bar q_p^i T^A u_r)\varepsilon_{ij}(\bar q_s^j T^A d_t)$ \,
  & & \\ \hline
	\end{tabular}
 	}
	\vspace{0.3cm}
	\\ \centering
	\scalebox{\OpScale}{%%%%%
	\setlength{\tabcolsep}{0.8mm}
    \hspace{1.4cm}
	\begin{tabular}{| rl | rl | rl | }
		\multicolumn{6}{c}{CP-conserving bosonic operators:~
    $9_{[\rm GIS]} \to 9_{[U(2)^5]}$~  
    (Table~\ref{tab:bosonic})} \\[.1cm] \hline
		\,
		$Q_{G}=$ & $f^{ABC} G_\mu^{A\nu} G_\nu^{B\rho} G_\rho^{C\mu}$ &	
		$Q_{H}=$ & $(H^\dagger H)^3$ &
		$Q_{HG}=$ & $(H^\dagger H) G_{\mu\nu}^A G^{A\mu\nu}$ 
        \\
       $Q_{W}=$ & $\varepsilon^{IJK} W_\mu^{I\nu} W_\nu^{J\rho} W_\rho^{K\mu}$ \,  &  \,
		$Q_{H\Box}=$ & $(H^\dagger H) \Box (H^\dagger H)$ & \,
        $Q_{HW}=$ & $(H^\dagger H) W_{\mu\nu}^I W^{I\mu\nu}$ 
        \\
		& &	
		$Q_{HD}=$ & $(H^\dagger D_\mu H)^\ast (H^\dagger D^\mu H)$ \, &
  		$Q_{H B}=$ & $(H^\dagger H) B_{\mu\nu} B^{\mu\nu}$ 
        \, \\
        & &	& & \,
		$Q_{H W B}=$ & $(H^\dagger \tau^I H) W_{\mu\nu}^I B^{\mu\nu}$ \,  \\
		\hline
	\end{tabular}
	}
  	\vspace{0.3cm}
 } %% end of small  
	\caption{List of all the independent  dimension-six SMEFT operators
    (excluding CP-odd terms). The operators are defined as in Ref.~\cite{Grzadkowski:2010es}. The labels $\{p,r,s,t\}$ denote flavor 
    indices. In each sector we indicate the number of independent gauge-invariant 
    structures (GIS), the number of independent operators in the 
    exact $U(2)^5$ limit, and the tables where we report current bounds on the corresponding WCs. 
	\label{tab:Warsaw-basis}
}
\end{table}
% \restoregeometry

\subsection{The flavor rotation}
\label{sec:flavor_rotation}

To proceed further, we need to specify the orientation of the $U(2)^5$ singlet fields in flavor space. In other words, we need to define what we identify as the third generation in higher-dimensional operators.  For right-handed fields, a non-ambiguous choice is dictated by the 
SM Yukawa couplings:
\begin{equation}
- \cL_{\rm SM}^{Y} = \sum_{r,p = 1,2,3} \bar \ell_r Y^{rp}_{e} e_p H +  \bar q_r Y^{rp}_{d} d_p H +  \bar q_r Y_{u}^{rp} u_p H_c + 
~{\rm h.c.} \,.
\label{eq:SMY}
\end{equation}
 In fact,  choosing the flavor basis where the $Y_f^\dagger Y_f$ matrices are diagonal ($f=u,d,e$), 
we  can  identify  the  right-handed
third-generation fields  as  those 
associated to the largest eigenvalues.
Neglecting neutrino masses, a similar non-ambiguous choice is possible also for the lepton doublets, 
since we can choose a flavor basis where  
$Y_e^\dagger Y_e$ and $Y_e Y_e^\dagger$ are both diagonal.

The only non-trivial case is related to the quark doublets, given the misalignment implied by the Cabibbo–Kobayashi–Maskawa (CKM) matrix ($V_{\rm CKM}$). 
Denoting as $q^{(u)}\equiv\{q_u, q_c, q_t\}$ and 
$q^{(d)}\equiv\{q_d, q_s, q_b\}$ the 
doublets in the up- and 
down-quark mass-eigenstate basis, respectively, 
we define the $U(2)_q$ singlet field as 
\begin{eqnarray}
q_3 &=& \big[ (1-\epsf) \delta_{3r} + \epsf V_{3r} \big] q_r^{(d)} \approx 
q_b +  \epsf (V_{ts} q_s + V_{td} q_d )  
\label{eq:epsf1} \\
&=&\big[ (1-\epsf) (V^\dagger)_{3r} + \epsf \delta_{3r} \big] q_r^{(u)} 
\approx  q_t + (1-\epsf)(V_{cb}^* q_c + V_{ub}^* q_u )\,. 
\label{eq:epsf2}
\end{eqnarray}
The approximate expressions on the r.h.s.~of Eqs.~(\ref{eq:epsf1})--(\ref{eq:epsf2}) hold in the limit $V_{tb}\to 1$, which is an excellent numerical approximation.
The free parameter $\epsf$, that we assume to be real and vary between 0 and 1 in the phenomenological analysis,  interpolates between  the down-aligned limit for the third generation 
($\epsf\to0$, ${q_3\to q_b}$) and the up-aligned limit ($\epsf\to1$, ${q_3 \to q_t}$), where 
\begin{equation}
q_b =\begin{pmatrix}  
    V^*_{ub} u_L + V^*_{cb} c_L + V^*_{tb} t_L \\
    b_L
\end{pmatrix}\,, \qquad 
 q_t =\begin{pmatrix}  
     t_L \\
    V_{td} d_L + V_{ts} s_L + V_{tb} b_L
\end{pmatrix}\,, 
\end{equation}
and the up- and down-type quark fields appearing on the right-hand-side of both equalities are defined in the mass basis. 

An equivalent way to describe this effect, and to better justify 
 Eqs.~(\ref{eq:epsf1})--(\ref{eq:epsf2}), is to consider the  
rotation matrices $L_{u,d}$  diagonalizing $Y_u Y_u^\dagger$ and  $Y_d Y_d^\dagger$ in the basis  where the SMEFT operators are  exactly invariant under $U(2)_q$. 
We define $L_d$ such that $L_d^\dagger (Y_d Y_d^\dagger) L_d  =  {\rm diag}(y_b^2, y_s^2, y^2_d)$, and similarly for $L_u$. By construction, $L_u$ and $L_d$ are not independent, since their product yields the CKM matrix: 
$V_{\rm CKM} = L_u^\dagger L_d$.
The  parametric form of $L_d$ (and $L_u$) can be determined 
under the assumption  of a minimal breaking of the $U(2)^5$ symmetry~\cite{Barbieri:2011ci}. This corresponds to assuming that the heavy-light mixing in the CKM matrix
is induced by a single spurion transforming as a {\bf 2} of $U(2)_q$.
Under this natural assumption, one obtains~\cite{Fuentes-Martin:2019mun}
\begin{equation}
 L_d \approx  \left(\begin{array}{c|c} 
 U_d &  {\begin{array}{c} 0  \\  \epsf V_{cb} \end{array}}  \\ \hline
 \epsf V_{td}  \quad   \epsf V_{ts}  & 1
 \end{array} \right)\,, 
 \qquad U_d = 
 \left(\begin{array}{cc} 
 c_d & -s_d e^{i\alpha} \\
 s_d e^{-i\alpha}  & c_d \\
 \end{array}
 \right)\,,
 \label{eq:Ld}
 \end{equation}
with $s_d/c_d = |V_{td} /V_{ts}|$ and $\alpha_d ={\rm arg}
(V_{td}^*/V^*_{ts})$.
This form for $L_d$ yields exactly Eqs.~(\ref{eq:epsf1})--(\ref{eq:epsf2}).  We emphasize that the $2\times 2$ unitary matrix $U_d$  has no impact in our analysis, given that we do not distinguish between the first two generations in the SMEFT. 
 
By means of the parameter $\epsf$, we effectively describe the leading $U(2)^5$ breaking in the quark sector.  
We do not consider dimension-six operators that explicitly violate the $U(2)^5$ symmetry, as could occur via direct spurion insertions in the SMEFT. However, diagonalizing the Yukawa couplings via the $L_{u,d}$ matrices, the $U(2)_q$-breaking spurion is ``transferred" from the Yukawa sector to the higher dimensional operators. As a result, by keeping $\epsf$ generic we can analyze the impact of this leading breaking in the whole SMEFT.

In principle, the $U(2)_q$-breaking spurion could appear in each operator with a different $O(1)$ coefficient; however, given that we analyze the bounds on the operators one-by-one, this makes no practical difference for terms with up to one spurion (which suffice to describe all $B$-physics observables).
The only cases where our effective description of the leading $U(2)_q$-breaking term gives rise to a difference compared to the most general description of this term is the correlation between heavy-light flavor-violating amplitudes  
($B$-physics) and light-light flavor-violating amplitudes ($D$- and $K$-physics) with the same
electroweak structure. In our approach, where flavor violation comes solely from rotations, this correlation is fixed, while in the more general case it is determined only up to  $O(1)$ coefficients.

%----------------------------------------------------------------------------
\section{Bounds from present data}
\label{sect:bounds}
%----------------------------------------------------------------------------

\subsection{Inputs and theory expressions}
\label{sect:inputs}

In order to constrain the $U(2)^5$ symmetric SMEFT, we combine bounds stemming from a plethora of experimental observations. We divide these into three macro categories: electroweak, flavor and collider 
observables.\footnote{Extensive  studies of 
one or more of these sectors, with particular focus on global 
fits of several SMEFT WCs, can be found in  Ref.~\cite{Ellis:2018gqa,DeBlas:2019ehy,Falkowski:2019hvp,Ethier:2021bye,Bellafronte:2023amz,Aoude:2020dwv,Bruggisser:2022rhb,Bruggisser:2021duo,Bissmann:2020mfi,Grunwald:2023nli}}
In this section  we discuss which observables are included in each category, their theory predictions, and the corresponding experimental measurements.

\paragraph{Electroweak observables.}
 In this category,  we include  $Z$- and $W$-pole measurements, lepton flavor universality (LFU) tests in $\tau$-decays, and flavor-conserving Higgs decays.
\begin{itemize}
\item As far as measurements at the $Z$- and $W$-pole are concerned, we include all the observables listed in Tables 1
and 2 of Ref. \cite{Breso-Pla:2021qoe}, using the SM predictions and experimental values given therein, as well as the same input scheme. We do not include the latest CDF II result on $m_W$ \cite{CDF:2022hxs}. We encode the NP contribution to these observables into a shift of the $W$ mass, $\delta m_W$, and into shifts of the $W$ and $Z$ couplings to SM fermions,  described by means of the $3 \times 3$ matrices $\delta g^{W \ell}$, $\delta g^{W q}$, $\delta g^{Z \, f}_L$ ($f=\,u,d,\nu, e$), and $\delta g^{Z f}_R$ ($f=\,u,d, e$), defined as in Appendix C.1 of Ref.~\cite{Allwicher:2023aql}.  The same reference contains expressions for $\delta m_W$ and $\delta g^{W,Z}$ in terms of SMEFT WCs, thus providing a map to express all pole observables in terms of these coefficients. 
\item We then consider tests of lepton flavor universality in leptonic and semi-leptonic $\tau$-decays. The relevant observables are $\left|g_\tau/g_\mu\right|$, $\left|g_\tau/g_e\right|$,  $\left|g_\tau/g_\mu\right|_\pi$, and $\left|g_\tau /g_\mu\right|_K$, defined via the ratios\footnote{We do not consider $\left|g_\mu/g_e\right|$, since  i) this measurement is not independent of the others and ii) it does not receive NP contributions in the limit of exact 
$U(2)^5$ symmetry.}
\begin{align}
   \begin{aligned}
  \left|\frac{g_\tau}{g_{\mu (e)}} \right| & = \frac{\mathcal{B}(\tau \to e (\mu) \nu \bar{\nu})/\mathcal{B}(\tau \to e (\mu) \nu \bar{\nu})_{\rm SM}}{\mathcal{B}(\mu \to e  \nu \bar{\nu})/\mathcal{B}(\mu \to e \nu \bar{\nu})_{\rm SM}} \,, \\
  \left|\frac{g_\tau}{g_\mu} \right|_{K (\pi)} & =  \frac{\mathcal{B}(\tau \to K(\pi) \nu)/\mathcal{B}(\tau \to K (\pi) \nu)_{\rm SM}}{\mathcal{B}(K  (\pi)\to \mu \bar \nu)/\mathcal{B}(K (\pi)\to \mu \bar \nu)_{\rm SM}} \,, 
  \label{eq:TauLFU_Definition}
   \end{aligned} 
\end{align}
which in the SM are 1 by construction. 
Theory expressions for these quantities in terms of Low-energy Effective Field Theory (LEFT) WCs at $\mu \sim m_\tau$ are given in~\cref{eq:TauLFUThVal}, while~\cref{eq:LnueMatching} provides the tree-level matching to the SMEFT  at the EW scale.  The corresponding experimental values and the correlation matrix are given in~\cref{eq:TauLFUExpVal} and~\cref{eq:TauLFUExpCorr}, as provided by HFLAV~\cite{HeavyFlavorAveragingGroup:2022wzx}.
\item The final set of observables in this category are the Higgs signal strengths, defined as 
\begin{equation}
\mu_{ii} = \frac{\mathcal{B}(h\to f_i \bar f_i)}{\mathcal{B}(h\to f_i \bar f_i)_{\rm SM}}\,,
\end{equation}
where again we have $\mu_{ii}^{\rm SM} =1$ by construction. The theory expression for the $\mu_{ii}$ in terms of SMEFT WCs is given in Eq. \eqref{eq:HiggsThVal}. Currently, only experimental measurements for $f = b,c,\tau,\mu$ are available \cite{ParticleDataGroup:2022pth}, which we report in~\cref{tab:HiggsExpVal}. 
\end{itemize}
\paragraph{Flavor observables.}
Under the category of flavor observables, we include flavor-changing $\Delta F = 1,2$ processes and the semi-leptonic charged-current transitions $b\rightarrow c(u)\ell\nu$. In particular, we have:
\begin{itemize}
    \item $\Delta F = 1$ flavor-changing neutral current (FCNC) observables. These include observables in the transitions $b \to s (d) \ell \ell$, $b (s)  \to s (d) \nu \bar{\nu}$, and $b \to s \gamma$. Regarding $b \to s (d) \ell \ell$, we include the branching ratios of the purely leptonic decays $B_{s(d)}  \to \mu^+ \mu^-$, as well as the branching ratios and angular observables of semileptonic $b \to s \ell \ell$ decays (e.g. $B \to K^{(\ast)} \ell \ell$, $B_s \to \phi \ell \ell$, etc.) . For $\mathcal{B}(B_{s(d)}  \to \mu^+ \mu^-)$ we take the updated experimental average given in Appendix C of~\cite{Greljo:2022jac} and use the SM theory prediction given in~\cite{Beneke:2019slt}. As to the remaining $b \to s \ell \ell$ measurements, we use the pseudo-observable $C_9^{\rm U}$ as a proxy for data. As our experimental value, we take $[C_9^{\rm U}]_{\rm exp} = -0.77 \pm 0.21$ as given in Appendix B of Ref.~\cite{Greljo:2022jac} (single parameter fit to $b\to s\mu\mu$ data). 
    All theory expressions in terms of SMEFT WCs are given in~\cref{app:bsllNP}. 
    
    Within $b(s) \rightarrow s(d) \nu\bar\nu$ transitions, we consider the branching ratios of $B^+ \to K^+ \nu \bar{\nu}$, $K^+ \to \pi^+ \nu \bar{\nu}$ and $K_L \to \pi_0 \nu \bar{\nu}$. For $B^+ \to K^+ \nu \bar{\nu}$ we use the average
    \begin{equation}
    \frac{\mathcal{B}(B^+ \rightarrow K^+ \nu\bar\nu)_{\rm exp}}{\mathcal{B}(B^+ \rightarrow K^+ \nu\bar\nu)_{\rm SM}} = 2.8 \pm 0.8 \,,
    \end{equation}
    which combines the recent experimental analysis of Belle-II~\cite{Glazov} with previous searches of this mode (as reported in~\cite{Glazov}), using the SM prediction in~\cite{Becirevic:2023aov}. For the analogous processes in $s \to d$ transitions, we use the experimental measurements given in~\cite{NA62:2021zjw,KOTO:2020prk} and the SM predictions in~\cite{Buras:2015qea}. The theory expressions for all these observables in terms of SMEFT WCs are given in \cref{app:bsnunuNP}.

    The last $\Delta F = 1$ observable we consider is the branching ratio of the radiative decay $B \to X_s \gamma$. For the experimental value, we use the current world average of $\mathcal{B}_{s\gamma}^{\rm exp} = (3.32 \pm 0.15)\times 10^{-4}$~\cite{HFLAV:2019otj}, while for the SM and NP theory contributions we closely follow~\cite{Misiak:2020vlo}. The map between the low-energy WCs entering this expression and the LEFT WCs is given in \cref{app:bsgammaNP}. As will be discussed later in more detail, using SMEFT-to-LEFT matching at 1-loop accuracy is crucial for this observable. We implement it using the expressions provided in the \texttt{Mathematica} notebook attached to the arxiv preprint of Ref.~\cite{Dekens:2019ept} as ancillary material.
    \item $\Delta F = 2$ observables. These include $B_{s,d}$ -, $D$-, and $K$-meson mixing. As pseudo-observable proxies for the relevant experimental measurements ($\epsilon_K$, $\Delta m_{B_{d,s}}$, ...), we use the bounds on the relevant low-energy WCs from~\cite{utfit}, which we report in~\cref{tab:DF2expbounds}. The  expressions of these quantities in terms of SMEFT WCs are given in~\cref{app:DF2}.
    \item Semi-leptonic charged-current processes involving the transitions $b \to c\ell \nu$ and $b \to u \ell \nu$. In this category we include the LFU ratios $R_D$ and $R_D^{\ast}$, defined as
    \begin{equation}
     R_{D^{(\ast)}}    = \frac{\mathcal{B}(B \to D^{(\ast)}  \tau \nu)}{\mathcal{B}(B \to D^{(\ast)}  \ell \nu)}\,,
    \end{equation}
    as well as the branching ratios 
    \begin{equation}
    \mathcal{B}(B_c \to \tau \nu) \qquad {\rm and} \qquad  \mathcal{B}(B_u \to \tau \nu) \,.
    \end{equation}
    For $R_{D^{(\ast)}}$, we take the values provided by HFLAV both for the experimental average as well as for the SM predictions~\cite{HFLAV:2019otj}, and we collect the relevant expressions in terms of SMEFT WCs in~\cref{app:FCCC}.
    For the $B_u$ branching fraction, we use 
    $\mathcal{B}\left(B_u \rightarrow \tau \nu \right)_{\mathrm{exp}}=1.09(24) \times 10^{-4}$~\cite{ParticleDataGroup:2022pth}
    and $\mathcal{B}\left(B_u \rightarrow \tau \nu \right)_{\mathrm{SM}}=0.812(54) \times 10^{-4}$~\cite{Bona:2022zhn}.
    For the $B_c$ case, only an upper bound is available. Following~\cite{Alonso:2016oyd,Akeroyd:2017mhr}, we take it to be $\mathcal{B}\left(B_c \rightarrow \tau \nu \right)_{\mathrm{exp}} < 0.1~(95\%~{\rm CL})$. For the SM theory prediction we use $\mathcal{B}\left(B_c \rightarrow \tau \nu \right)_{\mathrm{SM}}=0.023$~
    \cite{Alonso:2016oyd}. 
    Again, the NP theory expressions in terms of the SMEFT WCs are given in~\cref{app:FCCC}.
\end{itemize} 
\paragraph{Collider observables.}
This category includes measurements coming from $pp$ and $e^+e^-$ collisions above the $Z$ pole. In particular, we consider:
\begin{itemize}
    \item High-$p_T$ Drell-Yan tails, including LHC Run-II di- and mono-lepton data for all lepton flavors. We construct the relevant likelihood using \texttt{HighPT} ~\cite{Allwicher:2022mcg}. Since running effects for these observables are negligible, we ignore them.
    \item LEP-II data for $e^+e^- \to \ell^+\ell^-$. Following the same procedure as in \cite{Allanach:2023uxz}, we use the cross-section and forward-backward asymmetry measurements for $e^+e^-\to\mu^+\mu^-$ and $e^+e^-\to\tau^+\tau^-$, as well as the differential $e^+e^-\to e^+e^-$ data reported in \cite{ALEPH:2013dgf}. RG evolution is taken into account from each bin's energy up to 
     the reference scale $\Lambda_{\rm NP}=3$~TeV (see below).   
    \item  Jet observables. Four-quark contact interactions are constrained 
    using the results on the relevant combinations of SMEFT WCs provided in~\cite{Ethier:2021bye} (for four-quark operators with $q_L$,$t_R$) and \cite{Hartland:2019bjb} (for operators with $b_R$, including dipoles) using LHC data. Also in this case running effects are irrelevant.    
\end{itemize}

\subsection{Numerical analysis}
\label{sec:statAnal}
The first step needed to build our likelihood is to write the theory prediction for each observable in terms of SMEFT WCs at a reference high scale  $\Lambda_{\rm NP}$, which we take to be  $3$ TeV.  For observables with characteristic scale below the EW scale, this entails running the LEFT WCs entering the observable up
to the EW scale, matching to the SMEFT, and then running in the SMEFT up to $\Lambda_{\rm NP}$.  For LEFT running, LEFT-to-SMEFT matching, and SMEFT running, we employ \texttt{DSixTools} \cite{Celis:2017hod,Fuentes-Martin:2020zaz}, which enables us to work analytically in the WCs also beyond the leading-logarithmic (LL) approximation. Running effects within the LEFT are particularly significant for the dipole operators $L_{d \gamma}$ and $L_{d G}$ entering $B \to X_s \gamma$, as well as for the scalar and tensor operators $L_{\nu e d u}^{S, RL}$, $L_{\nu e d u}^{S, RR}$ and $L_{\nu e d u}^{T, RR}$, which enter observables in the charged-current $b \to c (u) \ell \nu$ transitions. 
 We consistently use tree-level SMEFT-to-LEFT matching conditions, except for operators contributing to $B \to X_s \gamma$ where, given the relevance of the constraint and the size of loop corrections, we implement one-loop matching conditions following~\cite{Dekens:2019ept} (see \cref{sect:results} for more details). 

Once all observables have been expressed in terms of SMEFT WCs at the scale $\Lambda_{\rm NP}$, we impose the $U(2)^5$ symmetry.\footnote{Note that this approach consistently allows for the breaking of $U(2)^5$ by RGE.} All theory expressions now depend on the 124  WCs  of the $U(2)^5$-symmetric SMEFT, which we collectively denote as $\vec{\mathcal{C}}(\Lambda_{\rm NP})$. At this stage, we construct the likelihood for each sector $\mathcal{S} = \{$EW, flavor, collider$\}$ as 
 \begin{equation}
\chi^2_{\mathcal{S}}(\vec{\mathcal{C}}(\Lambda_{\rm NP})) = \sum_{ij \in \mathcal{S}} \left[\mathcal{O}_i^{\rm th}(\vec{\mathcal{C}}(\Lambda_{\rm NP})) -\mathcal{O}_i^{\rm exp}\right] \sigma^{-2}_{ij} \left[\mathcal{O}_j^{\rm th}(\vec{\mathcal{C}}(\Lambda_{\rm NP}))-\mathcal{O}_j^{\rm exp}\right]\,, 
 \end{equation}
 where $\mathcal{O}_i^{\rm th}(\vec{\mathcal{C}}(\Lambda_{\rm NP}))$ is the theory prediction for the observable $\mathcal{O}_i$ written in terms of the high-scale $U(2)^5$-symmetric Wilson coefficients, $\mathcal{O}_i^{\rm exp}$ is the corresponding experimental value and $\sigma_{ij}$ is the covariance matrix for $\mathcal{O}_i$ and $\mathcal{O}_j$.
The global likelihood is then given by the sum of the individual likelihoods:
\begin{equation}
    \chi^2_{\rm tot} =     \chi^2_{\rm EW} +     \chi^2_{\rm flavor} +     \chi^2_{\rm collider} \,. 
\end{equation}
Subsequently, we switch on one coefficient $\mathcal{C}_k = \mathcal{C}_k(\Lambda_{\rm NP})$ at a time and determine the $3 \sigma$ confidence interval by requiring 
\begin{equation}
\chi^2_{\rm tot}(\mathcal{C}_k) - \chi^2_{\rm min} = 9\,.
\end{equation}
If more than two solutions exist for this equation, we select the two with the smallest absolute value. These are the solutions that would be obtained by using expressions linear in the WCs for the observables. We then convert this interval, denoted as  $[\mathcal{C}_k^{-},\mathcal{C}_k^{+}]$, into a range of scales $ [\Lambda_{-},\Lambda_+]$, where $\Lambda^{\pm} \equiv  (\mathcal{C}_k^{\pm})^{-1/2}$.  We then define $\Lambda_{\rm bound}$ as the minimum of $|\Lambda_{-}|$ and $|\Lambda_{+}|$, corresponding to the weakest bound.  

\subsection{Results}
\label{sect:results}
We present our results categorized by operator class in Tables~\ref{tab:psi2H2D}-\ref{tab:bosonic}. For each operator, we report the constraint on the corresponding WC at the scale $\Lambda_{\rm NP} = 3$ TeV from our three sectors -- flavor, EW, and collider -- as classified in \cref{sec:statAnal}. Because flavor bounds depend on the alignment choice, we provide the bounds both in the down-quark mass basis ($\epsf = 0$) as well as in the up-quark mass basis ($\epsf = 1$). The corresponding effective scales  are reported in the columns denoted by $\Lambda_{\rm flav.}^{\rm down}$ and $\Lambda_{\rm flav.}^{\rm up}$. The global bounds obtained by combining the likelihoods from all three sectors are given in the $\Lambda_{\rm all}^{\rm down}$ and $\Lambda_{\rm all}^{\rm up}$ columns, together with the observable that dominates each bound. In \cref{app:C} we report the bounds
from each sector separately, for both signs for the WC, indicating the most constraining observable in each direction.

%-------------------------------------
\begin{table}[t]
\centering
\small{
\renewcommand{\arraystretch}{1.3} 
\begin{tabular}{c|c|c|c|c|c|c|c|c}
coeff. & $\Lambda_\mathrm{flav.}^{\mathrm{down}}$ & $\Lambda_\mathrm{flav.}^{\mathrm{up}}$ & $\Lambda_\mathrm{EW}$ & $\Lambda_\mathrm{coll.}$ & $\Lambda_\mathrm{all}^{\mathrm{down}}$ & Obs. & $\Lambda_\mathrm{all}^{\mathrm{up}}$ & Obs.\\ \hline\hline
$\mathcal{C}_{H\ell }^{(1)[33]}$ & 0.1 & 0.1 & 4.4 & 1.6 & 4.3 & $R_\tau$ & 4.3 & $R_\tau$ \\
$\mathcal{C}_{H\ell }^{(1)[ii]}$ & 0.7 & 0.7 & 7.6 & 3. & 7.8 & $\sigma_{\rm had}$ & 7.8 & $\sigma_{\rm had}$ \\ \hline
$\mathcal{C}_{H\ell }^{(3)[33]}$ & 0.7 & 0.7 & 4.5 & 1.7 & 4.4 & $R_\tau$ & 4.4 & $R_\tau$ \\
$\mathcal{C}_{H\ell }^{(3)[ii]}$ & 0.7 & 0.7 & 7.7 & 3.8 & 7.7 & $\sigma_{\rm had}$ & 7.7 & $\sigma_{\rm had}$ \\ \hline
$\mathcal{C}_{He}^{[33]}$ & - & - & 3.8 & 1.5 & 3.7 & $R_\tau$ & 3.7 & $R_\tau$ \\
$\mathcal{C}_{He}^{[ii]}$ & 0.9 & 0.9 & 6.6 & 2.7 & 6.7 & $\sigma_{\rm had}$ & 6.7 & $\sigma_{\rm had}$ \\ \hline
$\mathcal{C}_{Hq}^{(1)[33]}$ & 0.3 & 5. & 3.7 & 0.1 & 3.7 & $\Gamma_Z$ & 5.1 & $B_s\to\mu\mu$ \\
$\mathcal{C}_{Hq}^{(1)[ii]}$ & 0.5 & 5.2 & 1.9 & 0.5 & 2. & $R_c$ & 5.4 & $B_s\to\mu\mu$ \\ \hline
$\mathcal{C}_{Hq}^{(3)[33]}$ & 1.3 & 5.6 & 3.5 & 0.4 & 3.4 & $R_b$ & 5.5 & $B_s\to\mu\mu$ \\
$\mathcal{C}_{Hq}^{(3)[ii]}$ & 1.3 & 5.3 & 5.6 & 3.1 & 5.7 & $R_\tau$ & 7.7 & $\Gamma_Z$ \\ \hline
$\mathcal{C}_{Hd}^{[33]}$ & - & - & 1.3 & 0.2 & 1.3 & $R_b$ & 1.3 & $R_b$ \\
$\mathcal{C}_{Hd}^{[ii]}$ & - & - & 1.7 & 0.3 & 1.7 & $R_\tau$ & 1.7 & $R_\tau$ \\ \hline
$\mathcal{C}_{Hu}^{[33]}$ & 0.6 & 0.6 & 3. & 0.1 & 3.1 & $A_b^{\rm FB}$ & 3.1 & $A_b^{\rm FB}$ \\
$\mathcal{C}_{Hu}^{[ii]}$ & - & - & 2.4 & 0.3 & 2.4 & $R_\tau$ & 2.4 & $R_\tau$ \\ \hline
\end{tabular}
}
\caption{Bounds on hermitian $\psi^2$ operators
(see text in \cref{sec:statAnal}--\ref{sect:results} for more details).}
\label{tab:psi2H2D}
\end{table}
%-------------------------------------
\begin{table}[t]
\centering
\small{
\renewcommand{\arraystretch}{1.3} 
\begin{tabular}{c|c|c|c|c|c|c|c|c}
coeff. & $\Lambda_\mathrm{flav.}^{\mathrm{down}}$ & $\Lambda_\mathrm{flav.}^{\mathrm{up}}$ & $\Lambda_\mathrm{EW}$ & $\Lambda_\mathrm{coll.}$ & $\Lambda_\mathrm{all}^{\mathrm{down}}$ & Obs. & $\Lambda_\mathrm{all}^{\mathrm{up}}$ & Obs.\\ \hline\hline
$\mathcal{C}_{eH}^{[33]}$ & - & - & 5.1 & - & 5.1 & $H\to \tau\tau$ & 5.1 & $H\to \tau\tau$ \\ \hline
$\mathcal{C}_{uH}^{[33]}$ & - & - & 0.2 & - & 0.2 & $H\to \tau\tau$ & 0.2 & $H\to \tau\tau$ \\ \hline
$\mathcal{C}_{dH}^{[33]}$ & - & - & 3.7 & - & 3.7 & $H\to bb$ & 3.7 & $H\to bb$ \\ \hline\hline
$\mathcal{C}_{Hud}^{[33]}$ & 3.2 & 3.2 & 0.5 & - & 3.2 & $B\to X_s \gamma$ & 3.2 & $B\to X_s \gamma$ \\ \hline\hline
$\mathcal{C}_{eB}^{[33]}$ & - & - & 0.2 & 1.2 & 1.2 & $pp\to\tau\tau$ & 1.2 & $pp\to\tau\tau$ \\ \hline
$\mathcal{C}_{uB}^{[33]}$ & 0.7 & 0.8 & 2.4 & 1.9 & 2.7 & $A_b^{\rm FB}$ & 2.7 & $A_b^{\rm FB}$ \\ \hline
$\mathcal{C}_{dB}^{[33]}$ & 15.2 & 74.8 & 0.4 & 0.7 & 15.2 & $B\to X_s \gamma$ & 74.8 & $B\to X_s \gamma$ \\ \hline
$\mathcal{C}_{eW}^{[33]}$ & - & - & 1. & 1.9 & 1.8 & $pp\to\tau\nu$ & 1.8 & $pp\to\tau\nu$ \\ \hline
$\mathcal{C}_{uW}^{[33]}$ & 0.5 & 0.9 & 2.3 & 3.6 & 3.7 & QuarkDipoles & 3.8 & QuarkDipoles \\ \hline
$\mathcal{C}_{dW}^{[33]}$ & 15.7 & 53. & 1.4 & 0.6 & 15.7 & $B\to X_s \gamma$ & 53. & $B\to X_s \gamma$ \\ \hline
$\mathcal{C}_{uG}^{[33]}$ & 0.1 & 0.3 & 0.5 & 2.7 & 2.7 & QuarkDipoles & 2.7 & QuarkDipoles \\ \hline
$\mathcal{C}_{dG}^{[33]}$ & 4. & 25.5 & 0.3 & - & 4. & $B\to X_s \gamma$ & 25.5 & $B\to X_s \gamma$ \\ \hline
\end{tabular}
}
\caption{Bounds on non-hermitian $\psi^2$ operators
(see text in \cref{sec:statAnal}--\ref{sect:results} for more details).}
\label{tab:dipole}
\end{table}
%-------------------------------------

The first notable observation is the complementarity between the three sectors. Out of the 124 bounds, 46 are dominated by EW observables, 42 by collider data, and 36 by flavor observables (assuming up-alignment, $\epsf = 1$).\footnote{In the down-aligned case, the numbers are 64 (EW), 45 (colliders) and 15 (flavor).} This emphasizes the need for a combined analysis encompassing all three sectors to obtain a comprehensive picture of the constraints from current data. This is especially important since EW and collider observables mainly constrain  flavor-conserving operators, while the flavor sector places stringent bounds on the flavor-violating ones. Regarding flavor-conserving operators, the strongest bounds from the EW sector are in the range $O(5-10)$ TeV for operators involving one or more Higgs fields and exhibit relatively weak dependence on the flavor of the fermions involved. As to collider data, the strongest bounds are again $O(5-10)$ TeV on four-fermion operators involving first-generation quarks and leptons, while operators with third-generation fermions receive milder constraints, typically 1 TeV.

While $U(2)^5$ is highly effective in reducing the scale associated with flavor violation, one still obtains bounds of $O(5-10)$ TeV on certain operators, for example four-fermion operators in the up-aligned scenario, akin to what happens in the MFV paradigm. However, a distinct advantage of $U(2)^5$ compared with MFV is that it gives the freedom to relax these bounds by allowing for some degree of alignment to the down-quark mass basis. A recurring theme in our results is that constraints on four-fermion operators involving third-generation fields can be eased to around a few TeV by down aligning, until the bounds become dominated by collider or EW observables. This implies that NP 
predominantly coupled to the third family at the TeV scale is consistent with bounds from all three sectors, an observation that we further investigate and quantify in~\cref{sect:third}.

\begin{figure}[t]
    \centering
    \begin{tikzpicture}[thick,>=stealth,scale=1]
        \node[wc] at (0,0) {};
        \node[below=7pt] at (0,0) {$\cC_{uu}$};
        \draw[midarrow] (0,0) -- (-1,1) node[left] {$u_R^3$};
        \draw[midarrow] (-1,-1) node[left] {$u_R^3$} -- (0,0);
        \draw[midarrow] (1,-1) -- (0,0);
        \draw[midarrow] (0,0) -- (1,1);
        \draw[midarrow] (1,1) -- (1,-1);
        \draw[dashed] (1,1) -- (2,1) node[right] {$H$};
        \draw[dashed] (1,-1) -- (2,-1) node[right] {$H$};
        \draw[->] (2.5,0) -- +(0.5,0) node[right] {$\cC_{Hu}$};
        \begin{scope}[xshift=7cm]
            \node[wc] at (0,0) {};
            \node[below=7pt] at (0,0) {$\cC_{Hu}$};
            \draw[dashed] (0,0) -- (-1,1) node[left] {$H$};
            \draw[dashed] (-1,-1) node[left] {$H$} -- (0,0);
            \draw[midarrow] (1,-1) -- (0,0);
            \draw[midarrow] (0,0) -- (1,1);
            \draw[midarrow] (1,1) -- (1,-1);
            \draw[dashed] (1,1) -- (2,1) node[right] {$H$};
            \draw[dashed] (1,-1) -- (2,-1) node[right] {$H$};
            \draw[->] (2.5,0) -- +(0.5,0) node[right] {$\cC_{HD}$};
        \end{scope}
    \end{tikzpicture}
    \caption{Next-to-leading log running of four-quark operators into $\mathcal{C}_{HD}$.
    \label{fig:RGEdiagrams}}
\end{figure}
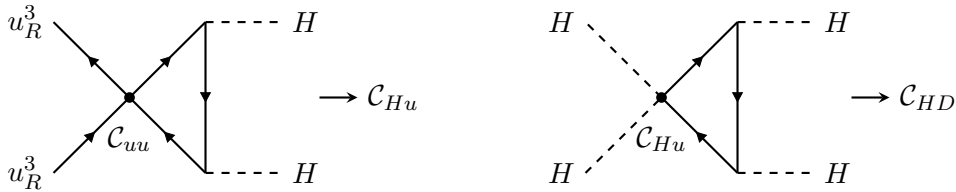
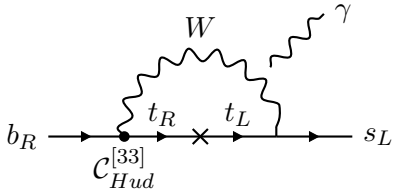
\begin{figure}[t]
    \centering
    \begin{tikzpicture}[thick,>=stealth,scale=1]
      \draw[vector] (-1,0) arc (180:0:1);
      \node[above=5pt] at (0,1) {$W$};
      \draw[midarrow] (-2,0) node[left] {$b_R$} -- (-1,0);
      \fill (-1,0) circle (2pt);
      \node[below] at (-1,0) {$\mathcal{C}_{Hud}^{[33]}$};
      \draw[midarrow] (-1,0) -- (0,0);
      \node[above] at (-0.5,0) {$t_R$};
      \draw[] (0,0) +(-0.1,-0.1) -- +(0.1,0.1);
      \draw[] (0,0) +(-0.1,0.1) -- +(0.1,-0.1);
      \node[above] at (0.5,0) {$t_L$};
      \draw[midarrow] (0,0) -- (1,0);
      \draw[midarrow] (1,0) -- (2,0) node[right] {$s_L$};
      \draw[vector] (45:1.3) -- (45:2.3) node[right] {$\gamma$};
    \end{tikzpicture}
    \caption{Finite one-loop matching contribution from $\cC_{Hud}^{[33]}$ to the photon dipole in the LEFT.
    \label{fig:bsgamma}}
\end{figure}

\subsubsection*{Importance of RGE  and one-loop matching}
We would like to emphasize a few crucial points  regarding the impact of 
RGE and matching conditions.
The first key point concerns the significance of RGE in the EW sector. 
Without considering RGE, only 16 of all the 124 $U(2)^5$-invariant operators affect $Z$- and $W$- pole observables. These are the bosonic operators $\cOp_{HWB}$ and $\cOp_{HD}$ (giving flavor-universal effects), the four-lepton operator $\cC_{\ell \ell}^{{ijji}}$ (that affects the extraction of $G_F$), and all operators in~\cref{tab:psi2H2D} except for $\cC_{Hu}^{[33]}$ (since there are no top quarks at the $Z$-pole). However, upon including RGE, 123 out of 124 operators are constrained by the EW sector (only $\cOp_H$ is not), and 44 of these operators receive bounds stronger than 1 TeV. The common feature among these 44 operators is that almost all of them contain top quarks, which yields substantial RG contributions to EW operators when closing the top loop and attaching Higgses. Interestingly, even though top quarks are not produced directly at the $Z$-pole, RG effects allow for competitive indirect bounds on top-quark operators via $Z$-pole observables, as recently highlighted in~\cite{Garosi:2023yxg}. For example, as illustrated in~\cref{tab:LLLL}, the indirect bounds on $\cC_{qq}^{(1,3)[3333]}$, $\cC_{lq}^{(1,3)[ii33]}$, and $\cC_{lq}^{(1,3)[3333]}$ from $Z$-pole observables are slightly stronger than the corresponding direct bounds from collider searches. Notably, the operator $\cOp_{Hu}^{[33]}$ (unconstrained by the EW sector at tree-level) obtains the most stringent constraint from the $Z$-pole observable $A_b^{\rm FB}$ due to its RG mixing into the flavor-universal $\cOp_{HD}$ (see Fig.~\ref{fig:RGEdiagrams}).

\begin{table}[p]
\centering
\small{
\renewcommand{\arraystretch}{1.3} 
\begin{tabular}{c|c|c|c|c|c|c|c|c}
coeff. & $\Lambda_\mathrm{flav.}^{\mathrm{down}}$ & $\Lambda_\mathrm{flav.}^{\mathrm{up}}$ & $\Lambda_\mathrm{EW}$ & $\Lambda_\mathrm{coll.}$ & $\Lambda_\mathrm{all}^{\mathrm{down}}$ & Obs. & $\Lambda_\mathrm{all}^{\mathrm{up}}$ & Obs.\\ \hline\hline
$\mathcal{C}_{\ell edq}^{[3333]}$ & 0.6 & - & 0.1 & 1.2 & 1.1 & $pp\to\tau\tau$ & 1.2 & $pp\to\tau\tau$ \\ \hline
$\mathcal{C}_{quqd}^{(1)[3333]}$ & 1.8 & 5.5 & 1.7 & 0.4 & 2.2 & $B\to X_s \gamma$ & 5.5 & $B\to X_s \gamma$ \\ \hline
$\mathcal{C}_{quqd}^{(8)[3333]}$ & 1. & 5.1 & 0.7 & 0.2 & 1. & $B\to X_s \gamma$ & 5.1 & $B\to X_s \gamma$ \\ \hline
$\mathcal{C}_{\ell equ}^{(1)[3333]}$ & - & - & 2.1 & - & 2.1 & $H\to \tau\tau$ & 2.1 & $H\to \tau\tau$ \\ \hline
$\mathcal{C}_{\ell equ}^{(3)[3333]}$ & - & - & 0.8 & - & 0.8 & $H\to \tau\tau$ & 0.8 & $H\to \tau\tau$ \\ \hline
\end{tabular}
}
\caption{Bounds on non-hermitian $\psi^4$ operators
(see text in \cref{sec:statAnal}--\ref{sect:results} for more details).}
\label{tab:nhpsi4}
\end{table}
%-------------------------------------
\begin{table}[p]
\centering
\small{
\renewcommand{\arraystretch}{1.3} 
\begin{tabular}{c|c|c|c|c|c|c|c|c}
coeff. & $\Lambda_\mathrm{flav.}^{\mathrm{down}}$ & $\Lambda_\mathrm{flav.}^{\mathrm{up}}$ & $\Lambda_\mathrm{EW}$ & $\Lambda_\mathrm{coll.}$ & $\Lambda_\mathrm{all}^{\mathrm{down}}$ & Obs. & $\Lambda_\mathrm{all}^{\mathrm{up}}$ & Obs.\\ \hline\hline
$\mathcal{C}_{\ell \ell }^{[3333]}$ & - & - & 0.3 & 0.2 & 0.3 & $\sigma_{\rm had}$ & 0.3 & $\sigma_{\rm had}$ \\
$\mathcal{C}_{\ell \ell }^{[ii33]}$ & - & - & 0.8 & 3.4 & 3.3 & $(e^+e^-\to\mu^+\mu^-)_{\rm FB}$ & 3.3 & $(e^+e^-\to\mu^+\mu^-)_{\rm FB}$ \\
$\mathcal{C}_{\ell \ell }^{[i33i]}$ & - & - & 3.3 & 3.3 & 4.2 & $(e^+e^-\to\mu^+\mu^-)_{\rm FB}$ & 4.2 & $(e^+e^-\to\mu^+\mu^-)_{\rm FB}$ \\
$\mathcal{C}_{\ell \ell }^{[iijj]}$ & - & - & 0.9 & 4.4 & 4.4 & $(e^+e^-\to\mu^+\mu^-)_{\rm FB}$ & 4.4 & $(e^+e^-\to\mu^+\mu^-)_{\rm FB}$ \\
$\mathcal{C}_{\ell \ell }^{[ijji]}$ & - & - & 4.5 & 4.4 & 4.9 & $A_b^{\rm FB}$ & 4.9 & $A_b^{\rm FB}$ \\ \hline
$\mathcal{C}_{qq}^{(1)[3333]}$ & 1. & 7.8 & 1.6 & 1.1 & 1.7 & $\Gamma_Z$ & 7.6 & $|C_{Bs}|$ \\
$\mathcal{C}_{qq}^{(1)[ii33]}$ & 1.3 & 11.2 & 0.9 & 1.5 & 1.7 & FourQuarksTop & 11.3 & $|C_{Bs}|$ \\
$\mathcal{C}_{qq}^{(1)[i33i]}$ & 2.5 & 11.3 & 0.7 & 1.6 & 2.6 & $B_s\to\mu\mu$ & 11.3 & $|C_{Bs}|$ \\
$\mathcal{C}_{qq}^{(1)[iijj]}$ & 0.9 & 8.1 & 0.4 & - & 0.9 & ${\rm Im}(C_D)$ & 8.1 & $|C_{Bs}|$ \\
$\mathcal{C}_{qq}^{(1)[ijji]}$ & 1.1 & 8.1 & 0.5 & - & 1. & ${\rm Im}(C_D)$ & 8.1 & $|C_{Bs}|$ \\ \hline
$\mathcal{C}_{qq}^{(3)[3333]}$ & 1. & 8.2 & 1.2 & 1.1 & 1.5 & $m_W$ & 8.2 & $|C_{Bs}|$ \\
$\mathcal{C}_{qq}^{(3)[ii33]}$ & 1.8 & 11.5 & 2.3 & 2.1 & 3. & $R_b$ & 11.3 & $|C_{Bs}|$ \\
$\mathcal{C}_{qq}^{(3)[i33i]}$ & 2.6 & 11.2 & 0.9 & 2.4 & 3.1 & $B_s\to\mu\mu$ & 11.3 & $|C_{Bs}|$ \\
$\mathcal{C}_{qq}^{(3)[iijj]}$ & 1. & 7.9 & 1.5 & 0.2 & 1.5 & $R_\tau$ & 7.9 & $|C_{Bs}|$ \\
$\mathcal{C}_{qq}^{(3)[ijji]}$ & 1.1 & 8. & 0.9 & 0.1 & 1.2 & $K^+\to\pi^+\nu\bar\nu$ & 8. & $|C_{Bs}|$ \\ \hline
$\mathcal{C}_{\ell q}^{(1)[3333]}$ & 0.1 & 1.7 & 1.4 & 1. & 1.4 & $R_\tau$ & 1.6 & $K^+\to\pi^+\nu\bar\nu$ \\
$\mathcal{C}_{\ell q}^{(1)[ii33]}$ & 0.4 & 5. & 2.5 & 1.5 & 2.5 & $\sigma_{\rm had}$ & 5.1 & $B_s\to\mu\mu$ \\
$\mathcal{C}_{\ell q}^{(1)[33ii]}$ & - & 1.6 & 0.3 & 3.4 & 3.4 & $pp\to\tau\tau$ & 3.4 & $pp\to\tau\tau$ \\
$\mathcal{C}_{\ell q}^{(1)[iijj]}$ & 0.5 & 5. & 0.5 & 5.4 & 5.4 & $pp\to\mu\mu$ & 5.6 & $pp\to\mu\mu$ \\ \hline
$\mathcal{C}_{\ell q}^{(3)[3333]}$ & 0.7 & 1.5 & 1.4 & 1. & 1.6 & $R_\tau$ & 1.6 & $K^+\to\pi^+\nu\bar\nu$ \\
$\mathcal{C}_{\ell q}^{(3)[ii33]}$ & 0.7 & 5.1 & 2.4 & 1.5 & 2.5 & $A_b^{\rm FB}$ & 5. & $B_s\to\mu\mu$ \\
$\mathcal{C}_{\ell q}^{(3)[33ii]}$ & 0.1 & 1.4 & 2. & 8.6 & 8.8 & $pp\to\tau\nu$ & 8.7 & $pp\to\tau\nu$ \\
$\mathcal{C}_{\ell q}^{(3)[iijj]}$ & 0.5 & 5.1 & 2.1 & 22.5 & 22.5 & $pp\to\mu\nu$ & 23.7 & $pp\to\mu\nu$ \\ \hline
\end{tabular}
}
\caption{Bounds on four-fermion $(\bar L L)(\bar L L)$ operators (see text in \cref{sec:statAnal}--\ref{sect:results} for more details).}
\label{tab:LLLL}
\end{table}
%-------------------------------------
\begin{table}[p]
\centering
\small{ 
\renewcommand{\arraystretch}{1.3} 
\begin{tabular}{c|c|c|c|c|c|c|c|c}
coeff. & $\Lambda_\mathrm{flav.}^{\mathrm{down}}$ & $\Lambda_\mathrm{flav.}^{\mathrm{up}}$ & $\Lambda_\mathrm{EW}$ & $\Lambda_\mathrm{coll.}$ & $\Lambda_\mathrm{all}^{\mathrm{down}}$ & Obs. & $\Lambda_\mathrm{all}^{\mathrm{up}}$ & Obs.\\ \hline\hline
$\mathcal{C}_{ee}^{[3333]}$ & - & - & 0.3 & 0.2 & 0.3 & $R_\tau$ & 0.3 & $R_\tau$ \\
$\mathcal{C}_{ee}^{[ii33]}$ & - & - & 0.7 & 3.2 & 3.2 & $(e^+e^-\to\mu^+\mu^-)_{\rm FB}$ & 3.2 & $(e^+e^-\to\mu^+\mu^-)_{\rm FB}$ \\
$\mathcal{C}_{ee}^{[iijj]}$ & - & - & 0.8 & 4.2 & 4.2 & $(e^+e^-\to\mu^+\mu^-)_{\rm FB}$ & 4.2 & $(e^+e^-\to\mu^+\mu^-)_{\rm FB}$ \\ \hline
$\mathcal{C}_{uu}^{[3333]}$ & 0.4 & 0.4 & 1.2 & 0.8 & 1.3 & $A_b^{\rm FB}$ & 1.3 & $A_b^{\rm FB}$ \\
$\mathcal{C}_{uu}^{[ii33]}$ & 0.1 & 0.1 & 1.1 & 1.3 & 1.4 & FourQuarksTop & 1.4 & FourQuarksTop \\
$\mathcal{C}_{uu}^{[i33i]}$ & - & - & 0.5 & 1.3 & 1.4 & FourQuarksTop & 1.4 & FourQuarksTop \\
$\mathcal{C}_{uu}^{[iijj]}$ & - & - & 0.3 & - & 0.3 & $R_\tau$ & 0.3 & $R_\tau$ \\
$\mathcal{C}_{uu}^{[ijji]}$ & - & - & 0.3 & - & 0.3 & $R_\tau$ & 0.3 & $R_\tau$ \\ \hline
$\mathcal{C}_{dd}^{[3333]}$ & - & - & - & - & - & $R_b$ & - & $R_b$ \\
$\mathcal{C}_{dd}^{[ii33]}$ & - & - & 0.1 & - & 0.1 & $R_\tau$ & 0.1 & $R_\tau$ \\
$\mathcal{C}_{dd}^{[i33i]}$ & - & - & - & - & - & $\Gamma_Z$ & - & $\Gamma_Z$ \\
$\mathcal{C}_{dd}^{[iijj]}$ & - & - & 0.2 & - & 0.2 & $R_\tau$ & 0.2 & $R_\tau$ \\
$\mathcal{C}_{dd}^{[ijji]}$ & - & - & 0.1 & - & 0.1 & $R_\tau$ & 0.1 & $R_\tau$ \\ \hline
$\mathcal{C}_{eu}^{[3333]}$ & - & - & 1.2 & 0.4 & 1.2 & $R_\tau$ & 1.2 & $R_\tau$ \\
$\mathcal{C}_{eu}^{[ii33]}$ & 0.9 & 0.9 & 2.1 & 0.7 & 2.2 & $\sigma_{\rm had}$ & 2.2 & $\sigma_{\rm had}$ \\
$\mathcal{C}_{eu}^{[33ii]}$ & - & - & 0.3 & 2.8 & 2.8 & $pp\to\tau\tau$ & 2.8 & $pp\to\tau\tau$ \\
$\mathcal{C}_{eu}^{[iijj]}$ & - & - & 0.6 & 7.4 & 7.4 & $pp\to ee$ & 7.4 & $pp\to ee$ \\ \hline
$\mathcal{C}_{ed}^{[3333]}$ & - & - & 0.2 & 1. & 1. & $pp\to\tau\tau$ & 1. & $pp\to\tau\tau$ \\
$\mathcal{C}_{ed}^{[ii33]}$ & - & - & 0.3 & 1.5 & 1.5 & $pp\to\mu\mu$ & 1.5 & $pp\to\mu\mu$ \\
$\mathcal{C}_{ed}^{[33ii]}$ & - & - & 0.2 & 2.8 & 2.8 & $pp\to\tau\tau$ & 2.8 & $pp\to\tau\tau$ \\
$\mathcal{C}_{ed}^{[iijj]}$ & - & - & 0.4 & 4.4 & 4.4 & $pp\to\mu\mu$ & 4.4 & $pp\to\mu\mu$ \\ \hline
$\mathcal{C}_{ud}^{(1)[3333]}$ & 0.1 & 0.1 & 0.4 & 0.3 & 0.4 & $R_b$ & 0.4 & $R_b$ \\
$\mathcal{C}_{ud}^{(1)[ii33]}$ & - & - & 0.1 & - & 0.1 & $R_\tau$ & 0.1 & $R_\tau$ \\
$\mathcal{C}_{ud}^{(1)[33ii]}$ & - & - & 0.5 & 1.2 & 1.2 & FourQuarksTop & 1.2 & FourQuarksTop \\
$\mathcal{C}_{ud}^{(1)[iijj]}$ & - & - & 0.2 & - & 0.2 & $R_\tau$ & 0.2 & $R_\tau$ \\ \hline
$\mathcal{C}_{ud}^{(8)[3333]}$ & 0.1 & 0.1 & - & 0.2 & 0.2 & FourQuarksBottom & 0.2 & FourQuarksBottom \\
$\mathcal{C}_{ud}^{(8)[ii33]}$ & - & - & - & - & - & - & - & - \\
$\mathcal{C}_{ud}^{(8)[33ii]}$ & - & - & 0.1 & 0.7 & 0.7 & FourQuarksTop & 0.7 & FourQuarksTop \\
$\mathcal{C}_{ud}^{(8)[iijj]}$ & - & - & - & - & - & - & - & - \\ \hline
\end{tabular}
}
\caption{Bounds on four-fermion $(\bar R R)(\bar R R)$ operators (see text in \cref{sec:statAnal}--\ref{sect:results} for more details).}
\label{tab:RRRR}
\end{table}
%-------------------------------------
\begin{table}[p]
\centering
\small{
\renewcommand{\arraystretch}{1.3} 
\begin{tabular}{c|c|c|c|c|c|c|c|c}
coeff. & $\Lambda_\mathrm{flav.}^{\mathrm{down}}$ & $\Lambda_\mathrm{flav.}^{\mathrm{up}}$ & $\Lambda_\mathrm{EW}$ & $\Lambda_\mathrm{coll.}$ & $\Lambda_\mathrm{all}^{\mathrm{down}}$ & Obs. & $\Lambda_\mathrm{all}^{\mathrm{up}}$ & Obs.\\ \hline\hline
$\mathcal{C}_{\ell e}^{[3333]}$ & - & - & 0.2 & 0.1 & 0.2 & $A_\tau$ & 0.2 & $A_\tau$ \\
$\mathcal{C}_{\ell e}^{[ii33]}$ & - & - & 0.4 & 2. & 1.9 & $(e^+e^-\to\mu^+\mu^-)_{\rm FB}$ & 1.9 & $(e^+e^-\to\mu^+\mu^-)_{\rm FB}$ \\
$\mathcal{C}_{\ell e}^{[33ii]}$ & - & - & 0.3 & 1.9 & 2. & $(e^+e^-\to\mu^+\mu^-)_{\rm FB}$ & 2. & $(e^+e^-\to\mu^+\mu^-)_{\rm FB}$ \\
$\mathcal{C}_{\ell e}^{[iijj]}$ & - & - & 0.5 & 3.8 & 3.8 & $(e^+e^-\to\mu^+\mu^-)_{\rm FB}$ & 3.8 & $(e^+e^-\to\mu^+\mu^-)_{\rm FB}$ \\ \hline
$\mathcal{C}_{\ell u}^{[3333]}$ & 0.1 & 0.1 & 1.4 & 0.4 & 1.3 & $R_\tau$ & 1.3 & $R_\tau$ \\
$\mathcal{C}_{\ell u}^{[ii33]}$ & 0.7 & 0.7 & 2.4 & 0.8 & 2.3 & $\sigma_{\rm had}$ & 2.3 & $\sigma_{\rm had}$ \\
$\mathcal{C}_{\ell u}^{[33ii]}$ & - & - & 0.4 & 3.1 & 3.1 & $pp\to\tau\tau$ & 3.1 & $pp\to\tau\tau$ \\
$\mathcal{C}_{\ell u}^{[iijj]}$ & - & - & 0.7 & 5.2 & 5.2 & $pp\to\mu\mu$ & 5.2 & $pp\to\mu\mu$ \\ \hline
$\mathcal{C}_{\ell d}^{[3333]}$ & - & - & 0.2 & 1. & 1. & $pp\to\tau\tau$ & 1. & $pp\to\tau\tau$ \\
$\mathcal{C}_{\ell d}^{[ii33]}$ & - & - & 0.3 & 1.5 & 1.5 & $pp\to\mu\mu$ & 1.5 & $pp\to\mu\mu$ \\
$\mathcal{C}_{\ell d}^{[33ii]}$ & - & - & 0.3 & 3. & 3. & $pp\to\tau\tau$ & 3. & $pp\to\tau\tau$ \\
$\mathcal{C}_{\ell d}^{[iijj]}$ & - & - & 0.5 & 4.7 & 4.7 & $pp\to\mu\mu$ & 4.7 & $pp\to\mu\mu$ \\ \hline
$\mathcal{C}_{qe}^{[3333]}$ & - & 0.3 & 1.2 & 1. & 1.3 & $R_\tau$ & 1.2 & $R_\tau$ \\
$\mathcal{C}_{qe}^{[33ii]}$ & 0.6 & 6.7 & 2.1 & 1.5 & 2.2 & $\sigma_{\rm had}$ & 6.7 & $B_s\to\mu\mu$ \\
$\mathcal{C}_{qe}^{[ii33]}$ & - & 0.3 & 0.2 & 3.7 & 3.7 & $pp\to\tau\tau$ & 3.7 & $pp\to\tau\tau$ \\
$\mathcal{C}_{qe}^{[iijj]}$ & - & - & 0.4 & 6. & 6. & $pp\to\mu\mu$ & 6. & $pp\to\mu\mu$ \\ \hline
$\mathcal{C}_{qu}^{(1)[3333]}$ & 0.3 & 1.8 & 1.2 & 0.6 & 1.3 & $\Gamma_Z$ & 1.7 & $B_s\to\mu\mu$ \\
$\mathcal{C}_{qu}^{(1)[ii33]}$ & 0.3 & 1.8 & 0.6 & 1.6 & 1.6 & FourQuarksTop & 2.1 & $B_s\to\mu\mu$ \\
$\mathcal{C}_{qu}^{(1)[33ii]}$ & - & 0.6 & 0.8 & 1.4 & 1.4 & FourQuarksTop & 1.2 & FourQuarksTop \\
$\mathcal{C}_{qu}^{(1)[iijj]}$ & - & 0.6 & 0.2 & - & 0.2 & $R_\tau$ & 0.6 & $|C_{Bd}|$ \\ \hline
$\mathcal{C}_{qu}^{(8)[3333]}$ & 0.2 & 0.7 & 0.1 & 0.4 & 0.4 & FourQuarksTop & 0.7 & $|C_{Bs}|$ \\
$\mathcal{C}_{qu}^{(8)[ii33]}$ & 0.3 & 0.7 & 0.1 & 1.2 & 1.2 & FourQuarksTop & 1.2 & FourQuarksTop \\
$\mathcal{C}_{qu}^{(8)[33ii]}$ & - & 0.1 & 0.2 & 0.8 & 0.8 & FourQuarksTop & 0.8 & FourQuarksTop \\
$\mathcal{C}_{qu}^{(8)[iijj]}$ & - & 0.1 & - & - & - & $R_\tau$ & 0.1 & $C_9^{\rm U}$ \\ \hline
$\mathcal{C}_{qd}^{(1)[3333]}$ & 0.2 & 0.3 & 0.4 & 0.3 & 0.3 & $R_b$ & 0.3 & $R_b$ \\
$\mathcal{C}_{qd}^{(1)[ii33]}$ & - & 0.3 & 0.1 & - & - & $R_\tau$ & 0.3 & $B_s\to\mu\mu$ \\
$\mathcal{C}_{qd}^{(1)[33ii]}$ & - & 0.4 & 0.6 & 1.3 & 1.2 & FourQuarksTop & 1.1 & FourQuarksTop \\
$\mathcal{C}_{qd}^{(1)[iijj]}$ & - & 0.4 & 0.2 & - & 0.2 & $R_\tau$ & 0.4 & $B_s\to\mu\mu$ \\ \hline
$\mathcal{C}_{qd}^{(8)[3333]}$ & - & - & - & 0.2 & 0.2 & FourQuarksBottom & 0.2 & FourQuarksBottom \\
$\mathcal{C}_{qd}^{(8)[ii33]}$ & 0.1 & - & - & - & 0.1 & $B\to X_s \gamma$ & - & $B\to X_s \gamma$ \\
$\mathcal{C}_{qd}^{(8)[33ii]}$ & - & - & 0.1 & 0.7 & 0.7 & FourQuarksTop & 0.7 & FourQuarksTop \\
$\mathcal{C}_{qd}^{(8)[iijj]}$ & - & - & - & - & - & $R_\tau$ & - & $|C_{Bs}|$ \\ \hline
\end{tabular}
}
\caption{Bounds on four-fermion $(\bar L L)(\bar R R)$ operators (see text in \cref{sec:statAnal}--\ref{sect:results} for more details).}
\label{tab:LLRR}
\end{table}
%-------------------------------------
\begin{table}[t]
\centering
\small{
\renewcommand{\arraystretch}{1.3} 
\begin{tabular}{c|c|c|c|c|c|c|c|c}
coeff. & $\Lambda_\mathrm{flav.}^{\mathrm{down}}$ & $\Lambda_\mathrm{flav.}^{\mathrm{up}}$ & $\Lambda_\mathrm{EW}$ & $\Lambda_\mathrm{coll.}$ & $\Lambda_\mathrm{all}^{\mathrm{down}}$ & Obs. & $\Lambda_\mathrm{all}^{\mathrm{up}}$ & Obs.\\ \hline\hline
$\mathcal{C}_{H}$ & - & - & - & - & - & - & - & - \\ \hline
$\mathcal{C}_{H\Box}$ & 0.2 & 0.2 & 0.6 & 0.1 & 0.6 & $A_b^{\rm FB}$ & 0.6 & $A_b^{\rm FB}$ \\ \hline
$\mathcal{C}_{HD}$ & 0.5 & 0.5 & 5.1 & - & 5. & $A_b^{\rm FB}$ & 5. & $A_b^{\rm FB}$ \\ \hline
$\mathcal{C}_{HG}$ & 0.8 & 0.8 & 0.4 & - & 0.9 & $B\to X_s \gamma$ & 0.9 & $B\to X_s \gamma$ \\ \hline
$\mathcal{C}_{HB}$ & 0.5 & 0.5 & 0.9 & - & 0.9 & $A_b^{\rm FB}$ & 0.9 & $A_b^{\rm FB}$ \\ \hline
$\mathcal{C}_{HW}$ & 0.7 & 0.7 & 0.9 & - & 1. & $A_b^{\rm FB}$ & 1. & $A_b^{\rm FB}$ \\ \hline
$\mathcal{C}_{HWB}$ & 1. & 1. & 9. & - & 9. & $A_b^{\rm FB}$ & 9. & $A_b^{\rm FB}$ \\ \hline
$\mathcal{C}_{G}$ & 1.1 & 1.1 & 0.1 & - & 1.1 & $B\to X_s \gamma$ & 1.1 & $B\to X_s \gamma$ \\ \hline
$\mathcal{C}_{W}$ & 0.3 & 0.3 & 0.9 & - & 0.9 & $A_b^{\rm FB}$ & 0.9 & $A_b^{\rm FB}$ \\ \hline
\end{tabular}
}
\caption{Bounds on CP-conserving bosonic operators
(see text in \cref{sec:statAnal}--\ref{sect:results} for more details).}
\label{tab:bosonic}
\end{table}

The second important point is the relevance of going beyond the LL approximation when solving the RG equations. Even though the scale hierarchy we deal with is not particularly large, next-to-leading log (NLL) effects yield corrections as large as $30\%$ to the LL results, which inadvertedly tend to relax the LL constraints. A particularly interesting case is that of $\cOp_{uu}^{[3333]}$, which enters the EW fit only at NLL by mixing into $\cOp_{HD}$, as illustrated in
Figure~\ref{fig:RGEdiagrams}. Although it is parameterically a two-loop effect, this NLL running can be sizeable, as  pointed out in~\cite{Allwicher:2023aql}, leading to a bound of 1.2 TeV on $\cC_{uu}^{[3333]}$. Surprisingly, this purely NLL running into a $Z$-pole observable results in a more stringent constraint on this operator than direct collider searches, which yield a bound of about 800 GeV. 

Lastly, it is worth stressing that in selected observables,
 one-loop matching from the SMEFT to the LEFT can be crucial in order to obtain the correct low-energy phenomenology. A clear example is provided by the operator $\cOp_{Hud}^{[33]}$, which leads to the effective vertex $W t_R b_R$ after EW symmetry breaking. Although the operator itself is flavor conserving, it can be inserted at one loop in order to give a NP contribution to $B \rightarrow X_s \gamma$, as shown in Figure~\ref{fig:bsgamma}. Furthermore, this NP contribution is $m_t$ enhanced, lifting the $m_b$ chiral suppression found in the SM and allowing TeV-scale NP to give SM-sized contributions. Notably, the diagram in Figure~\ref{fig:bsgamma} is finite, meaning that it cannot be estimated using RGE. In fact, it only appears as a finite contribution in the SMEFT-to-LEFT matching computation~\cite{Dekens:2019ept}. Through this finite contribution to $B\rightarrow X_s \gamma$, the flavor sector imposes by far the strongest bound on $\cC_{Hud}^{[33]}$ at 3.2 TeV, which is otherwise only weakly constrained by $y_b$-suppressed running into the EW sector.

%-----------------------------------------------------------------------------
\section{ The hypothesis of third-generation NP}
\label{sect:third}

%-----------------------------------------------------------------------------
\subsection{Motivation and definition of the setup}
\label{sect:thirdA}

As previously mentioned, the $U(2)^5$ symmetry significantly alleviates the bounds on the NP scale(s) compared with the generic SMEFT. However, several bounds still exceed 10 TeV, as shown in Tables~\ref{tab:psi2H2D}--\ref{tab:bosonic}. This section aims  to identify general conditions that can further reduce these bounds, with the overarching goal of investigating scenarios where NP couples mainly to the third generation. 

The motivation for exploring this hypothesis is twofold. First, models where NP is predominantly coupled to third-generation fermions have a strong theoretical motivation, such as addressing the EW hierarchy problem and accommodating the hierarchical structure of the Yukawa couplings. Second, some of the most stringent flavor-conserving constraints in the SMEFT arise from collider processes featuring only light quarks and leptons in the final state (as can be seen in the lower part of Table~\ref{tab:LLLL}). Hence, models with suppressed couplings to 
light families have the potential to remain compatible with exiting data, even for relatively low NP scales. 

It is important to notice that the $U(2)^5$ symmetry, on its own, does not provide any specific insight into the underlying dynamics, aside from indicating the absence of flavor violation involving the light families. To characterize the class of models we are interested in, we need to make further assumptions about the parametric structure of the WCs at the high scale $\Lambda_{\rm NP}$. We outline these assumptions as follows:
\begin{itemize} 
\item[i)] The WCs of operators involving light fields are subject to a suppression factor $(\epsq)^{n_Q} (\epsl)^{n_L}$, where $n_Q$ ($n_L$) denotes the number of light quark (lepton) fields in the operator. For example, we assume  $\cC_{qe}^{[iijj]}= \epsq^2 \epsl^2/\Lambda^2$.
\item[ii)] The WCs of operators containing one or more Higgs fields feature a suppression factor $(\epsh)^{n_H}$, where $n_H$ represents the number of Higgs fields. 
\item[iii)] The WCs of operators that include one or more gauge field strength tensors are suppressed by $\epsloop = (\Pi_i g_i)/(16\pi^2)$, where the product of gauge couplings runs over the field strength tensors present in the operator. For example, $\cC_{dW}^{[33]}=  \epsh g_L/(16\pi^2 \Lambda^2)$ 
and $\cC_{H B}= \epsh^2 g_Y^2/(16\pi^2 \Lambda^2)$. 
\end{itemize}
Only the first of these conditions is directly related to the hypothesis of NP coupled mainly to the third generation. The second condition serves to distinguish NP couplings with the Higgs from those with fermions.  
The last condition holds in a broader context, being applicable to all UV completions where operators with field strength tensors are generated beyond the tree level. 

Therefore, the only suppression factor that we assume \emph{a priori} is $\epsloop$.  For all other suppression factors $\varepsilon_i$, including the parameter $\epsf$ controlling the deviation from down alignment, our strategy is to use data to determine the maximum allowed values that consistently permit NP to reside at the TeV scale.

\subsection{Conditions for TeV-scale new physics}
\label{sect:thirdB}

As per the rules defined above, the only operators without  
any inherent suppression factor are the four-fermion operators involving solely 
third-generation fields. A close examination of the $\cC_i^{[3333]}$ entries in Tables~\ref{tab:nhpsi4}--\ref{tab:LLRR} reveals that the 
corresponding bounds typically fall around or below 
\begin{equation}
    \Lambda_0 = 1.5~{\rm TeV}\,,
\end{equation}
at least in the more favorable case of down 
alignment.\footnote{The sole (slight) exception to this trend
is $\mathcal{C}_{quqd}^{(1)[3333]}$, 
whose bound in the down-alignment limit is 2.2~TeV.}
As a consequence, in this class of models, going below this reference NP scale is infeasible in the absence of more specific, model-dependent assumptions. Investigating such assumptions goes beyond the scope of this work. Our goal is to determine whether the bounds on {\em all} other operators can be compatible with the reference scale $\Lambda_0$ by assuming reasonable values for the $\varepsilon_i$.

\begin{figure}[p]
    \centering
    \includegraphics[width=1.0\textwidth]{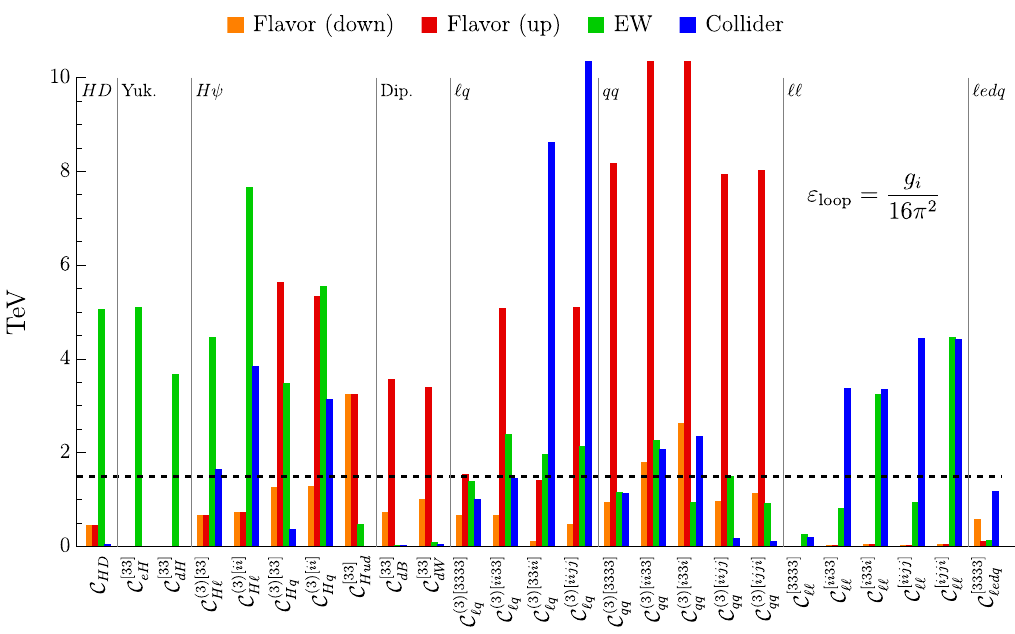}
    \caption{Bounds on the coefficients of a representative set of $d=6$ operators, expressed as bounds on the corresponding effective scale, after imposing the $\epsloop$ suppression factor on loop-generated terms (see Sect.~\ref{sect:thirdB}). }
    \label{fig:plot1}
\end{figure}

\begin{figure}[p]
    \centering
\includegraphics[width=1.0\textwidth]{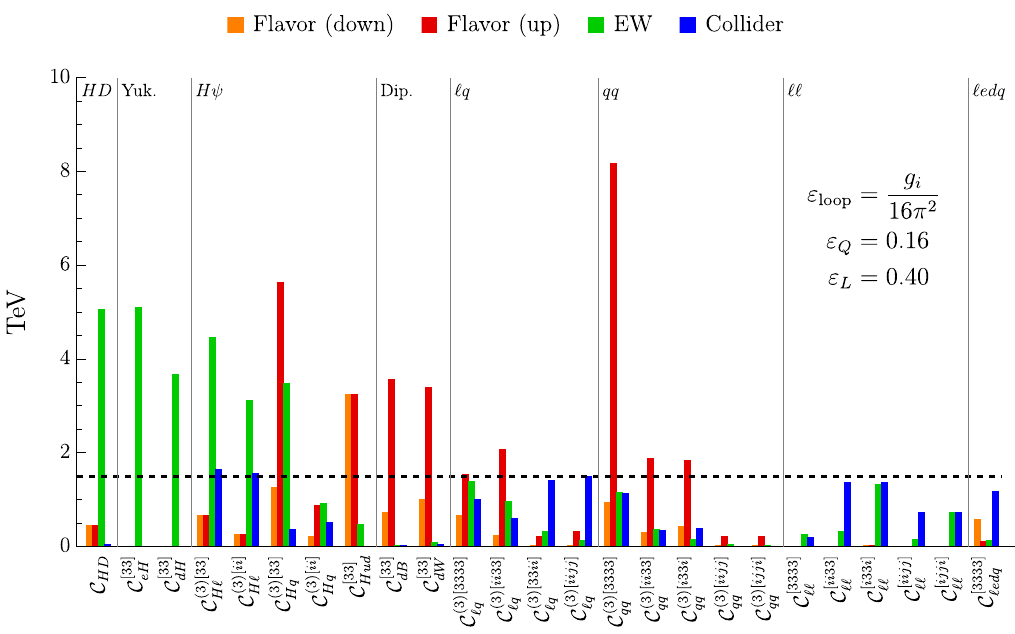}
    \caption{Bounds on the same operators shown in Fig.~\ref{fig:plot1} after imposing 
    the $\epsl$ and $\epsq$  suppression factors (see Sect.~\ref{sect:thirdB}).}
    \label{fig:plot2}
\end{figure}

\begin{figure}[p]
    \centering
    \includegraphics[width=1.0\textwidth]{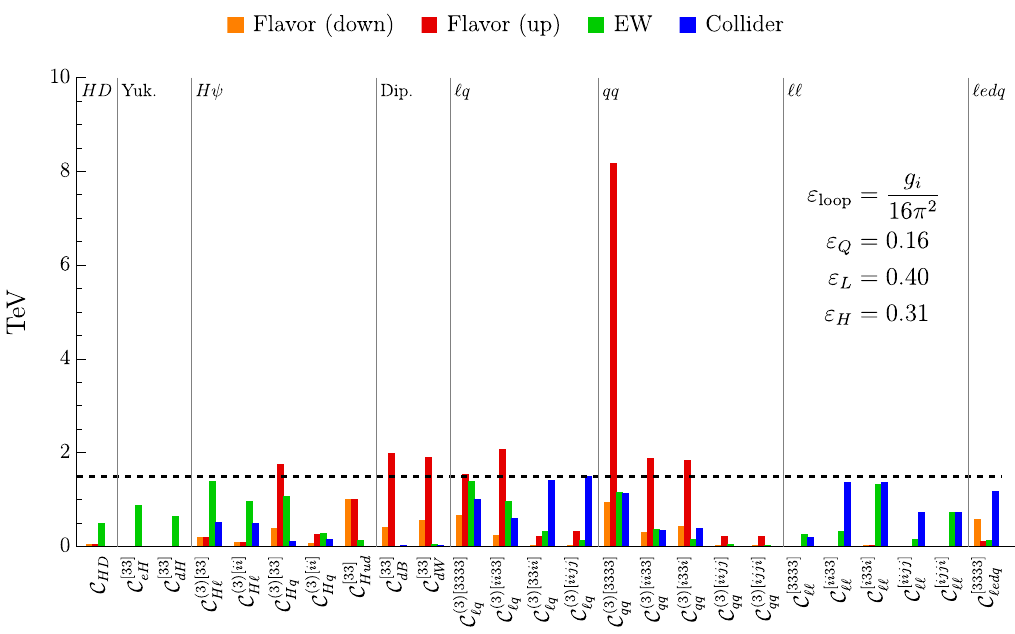}
    \caption{Bounds on the same operators shown in Fig.~\ref{fig:plot2} after imposing 
    the $\epsh$  suppression factor (see Sect.~\ref{sect:thirdB})}
    \label{fig:plot3}
\end{figure}

\begin{figure}[p]
    \centering
    \includegraphics[width=1.0\textwidth]{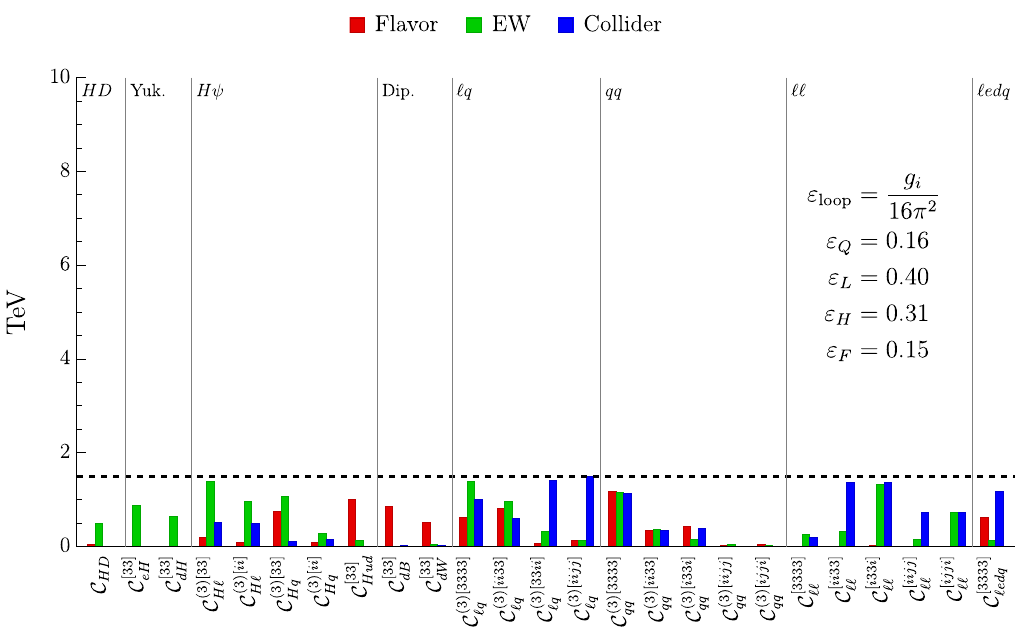}
    \caption{Bounds on the same operators shown in Fig.~\ref{fig:plot3} fixing $\epsf$ (see Sect.~\ref{sect:thirdB}).}
    \label{fig:plot4}
\end{figure}

To achieve this goal we proceed as follows. We first implement the \emph{a priori} determined suppression factor $\epsloop$. The resulting bounds for a representative set of operators are shown in Fig.~\ref{fig:plot1}.
The second step is determining upper bounds for $\epsq$  
and $\epsl$ by considering collider constraints on semi-leptonic  operators. Since these constraints are largely independent of flavor 
alignment, the resulting upper bounds on $\epsq$ and $\epsl$ are model-independent. The resulting reduction in NP bounds corresponds to the evolution  Fig.~\ref{fig:plot1}~$ \to $~Fig.~\ref{fig:plot2}. With $\epsq$ and $\epsl$ fixed,  the third step is to determine the upper bound on $\epsh$ using EW constraints on operators with Higgs fields (Fig.~\ref{fig:plot2}~$\to$~Fig.~\ref{fig:plot3}). 
Having fixed all suppression factors related to the operator structure, we can finally ask what is the maximal deviation from down alignment (i.e. which is the maximum allowed value of  $\epsf$)  for which the bounds from flavor-violating observables also do not exceed $\Lambda_0$. This yields the evolution Fig.~\ref{fig:plot3}~$\to$~Fig.~\ref{fig:plot4}.

Through this procedure, we deduce that under the parametric scaling of the WCs defined in Sect.~\ref{sect:thirdA},
if
\begin{equation}
    \epsq \leq 0.16\,, \qquad 
    \epsl \leq 0.40\,, \qquad  
    \epsh \leq 0.31\,, \qquad      
    \epsf \leq 0.15\,, 
    \label{eq:epsbound}
\end{equation}
then all dimension-six operators in our EFT are compatible with an effective scale as low as 1.5~TeV. 
It is important to note that there is no special meaning to 
the values of the $\varepsilon_i$ 
in Eq.~(\ref{eq:epsbound}). In realistic models, each operator will have  
a specific scaling with the couplings of the underlying theory.\footnote{To this purpose, an interesting indication is obtained by identifying the relative weights for the WCs  associated to specific tree-level mediators, as discussed  recently in~\cite{Greljo:2023adz} in the $U(3)^5$ symmetric case.}
However, these values provide a valuable semi-quantitative indication of the general UV conditions that NP models at nearby energy scales need to satisfy to be compatible with present data.

The most crucial point to emphasize from this analysis is that the upper bounds in Eq.~(\ref{eq:epsbound}) are not particularly stringent: they are all well above the size 
one would expect if the corresponding effective 
operators were generated beyond the tree level.
In other words, these constraints serve as perfectly valid UV matching 
conditions that will not be spoiled by radiative corrections. 
This implies that it is possible to envision realistic extensions of the SM involving NP at the TeV scale 
dominantly coupled to the third generation.

%-----------------------------------------------------------------------------
\section{Future projections}
\label{sect:future}
%-----------------------------------------------------------------------------
In this section, we investigate the impact of projected improvements in current flavor and collider experiments 
in the near future, as well as the implications of a future circular $e^+ e^-$ machine from a longer-term perspective.

\subsection{Interplay of flavor, collider, 
and EW data: A motivated example}

To illustrate the role of precision measurements in flavor physics in the near future,
we focus on the rare decays $B^+ \rightarrow K^+ \nu\bar\nu$ and $K^+ \rightarrow \pi^+ \nu\bar\nu$, exploring their interplay with related electroweak and collider observables. These rare processes hold particular interest for a series of reasons:
i) they are theoretically clean, making it straightforward to interpret possible deviations from the SM; ii) the measurements of the corresponding decay rates are expected to improve significantly in the next 10-15 years; iii) they are sensitive to a limited number of effective operators, allowing for a simple EFT interpretation;
iv) the present central values exceed the corresponding SM predictions (with different degrees of significance), allowing for speculation about possible  beyond-the-SM  contributions.

Using the experimental inputs and SM predictions described in Sect.~\ref{sect:inputs}, present experimental results, normalized to the corresponding SM predictions, are as follows: 
\begin{equation}
\frac{\mathcal{B}(B^+ \rightarrow K^+ \nu\bar\nu)_{\rm exp}}{\mathcal{B}(B^+ \rightarrow K^+ \nu\bar\nu)_{\rm SM}} = 2.8 \pm 0.8 \,,
\qquad 
\frac{\mathcal{B}(K^+ \rightarrow \pi^+ \nu\bar\nu)_{\rm exp}}{\mathcal{B}(K^+ \rightarrow \pi^+ \nu\bar\nu)_{\rm SM}} = 1.23 \pm 0.39 \,.
\end{equation}

In the $U(2)^5$-symmetric SMEFT, flavor violation is limited to left-handed quarks. The low-energy effective Lagrangian describing $b\rightarrow s\nu\bar \nu$ and $s\rightarrow d\nu\bar \nu$ transitions at the partonic level can be conveniently normalized as 
\begin{equation}
\mathcal{L}_{d_i \rightarrow d_j \nu\bar \nu} = \frac{4 G_F}{\sqrt{2}} 
 \frac{\alpha_W}{2\pi} \left[ 
 V_{ts}^* V_{tb}  C_L^{sb\ell} (\bar s_L \gamma_{\mu} b_L) + 
 V_{td}^* V_{ts}  C_L^{ds\ell} (\bar d_L \gamma_{\mu} s_L) 
 \right] \times 
  (\bar \nu_L^\ell \gamma^\mu \nu_L^\ell) 
 \,,
\end{equation}
where $4\pi \alpha_W = g_L^2$. Given the absence of QCD and QED contributions to the anomalous dimensions of these effective operators, we can determine possible NP contributions via a simple tree-level matching to the SMEFT. Decomposing the $C_L^{ij\ell}$ as $C_L^{ij\ell} \equiv C_{L, \rm SM}^{ij\ell} + \Delta C_L^{ij\ell}$, this leads to
\begin{align}
    \Delta C_{L}^{ij \ell} &=  \frac{\pi v^2}{\alpha_W V_{ti}^* V_{tj}} [L_{\nu d}^{V,LL}]_{\ell \ell ij} \,,
\end{align}    
where
\begin{align}    
    [L_{\nu d}^{V,LL}]_{\ell \ell ij} = [L_d^\dagger]_{i p} \Big[ \cC_{\ell q}^{(1)[\ell\ell pr]} - \cC_{\ell q}^{(3)[\ell\ell pr]} + \cC_{Hq}^{(1)[pr]} + \cC_{Hq}^{(3)[pr]} \Big] [L_d]_{rj}   \,.
    \label{eq:LVLL}
\end{align}
The following considerations allow us to restrict our attention to a specific subset of the $\cC_i$ appearing in Eq.~(\ref{eq:LVLL}):
\begin{itemize}
    \item The combination 
    $\cC_{Hq}^{(1)[pr]} + \cC_{Hq}^{(3)[pr]}$  in
    Eq.~(\ref{eq:LVLL}) is tightly constrained by $\mathcal{B}(B_{s} \rightarrow \mu^+ \mu^-)$. 
   \item Both $\cC_{\ell q}^{(1)}$ and $\cC_{\ell q}^{(3)}$ with light quarks face strong constraints from Drell-Yan processes.
    \item  $\cC_{\ell q}^{(1)[ii33]}$ and $\cC_{\ell q}^{(3)[ii33]}$ are tightly constrained by $\mathcal{B}(B_{s} \rightarrow \mu^+ \mu^-)$ and, to a smaller extent, by EWPO.   
\end{itemize} 
Given these considerations,
we conclude that, in our framework, the only possibility to generate sizable deviations from the SM in these rare neutrino modes is via $\cC_{\ell q}^{(1)[3333]}$ and $\cC_{\ell q}^{(3)[3333]}$.
Remarkably, these are also 
the only non-suppressed WCs in the general class of NP extensions described in Sect.~\ref{sect:third}.

\begin{figure}[t]
    \centering
    \includegraphics[width=0.4675\textwidth]{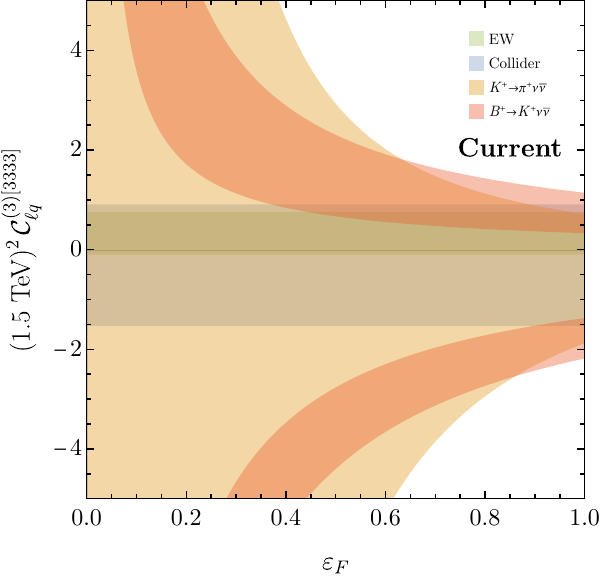} \hspace{7.55mm}
    \includegraphics[width=0.45\textwidth]{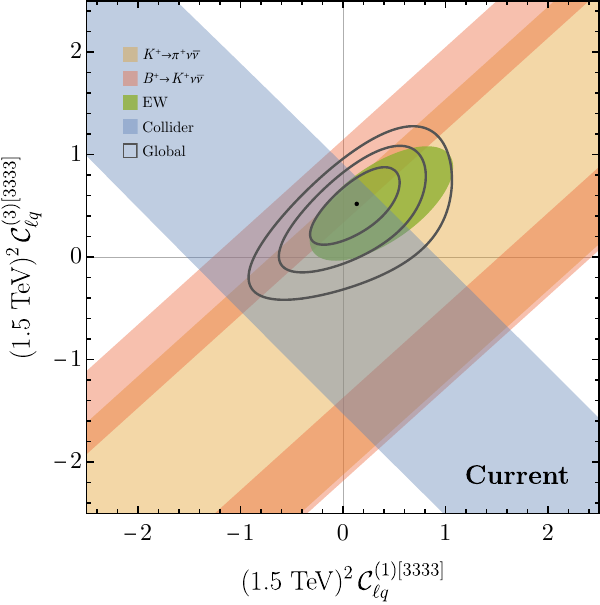} \hspace{7.5mm}
    \includegraphics[width=0.4675\textwidth]{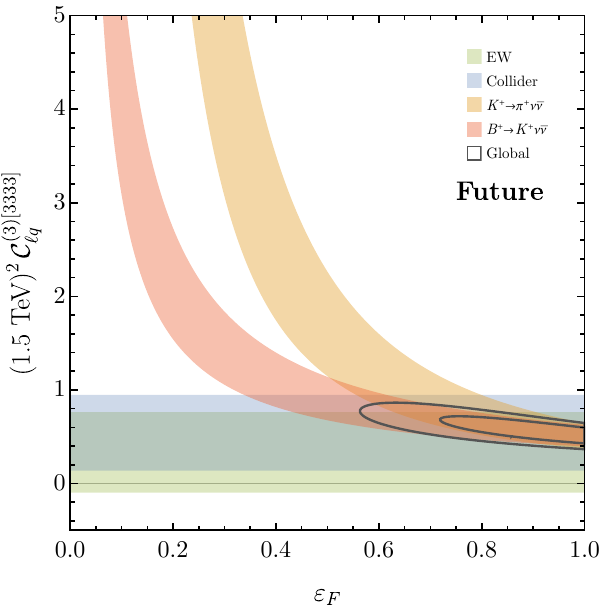} \hspace{7.55mm}
    \includegraphics[width=0.465\textwidth]{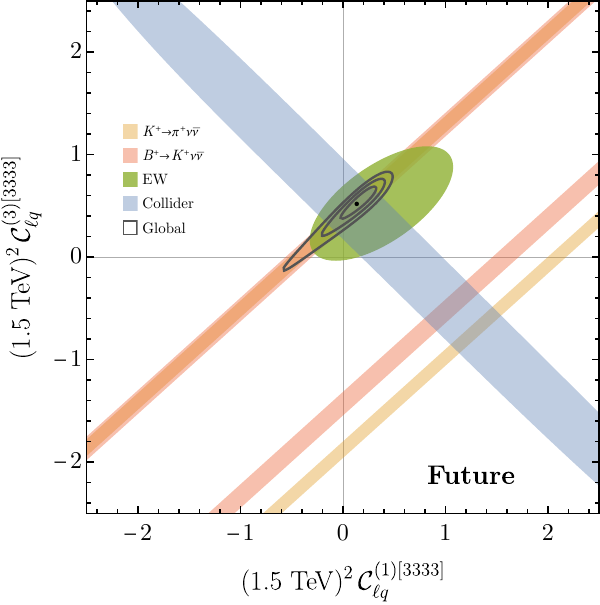} \hspace{7.5mm}
    \caption{Upper-left plot: current constraints from      $\cB(B^+ \rightarrow K^+ \nu\bar\nu)$,
    $\cB(K^+ \rightarrow \pi^+ \nu\bar\nu)$, EWPO 
    and collider observables on the Wilson coefficient 
    $\cC_{\ell q}^{(3)[3333]}$ (with all other WCs set to zero) as a function of the flavor-alignment parameter $\epsf$. Upper-right plot: constraints from the same set of observables in the $\cC_{\ell q}^{(1)[3333]}$--$\cC_{\ell q}^{(3)[3333]}$ plane, setting $\epsf=1$.
    Lower plots: possible evolution of the corresponding upper plots using future projections for flavor and colliders observables  (see text).
    In all plots the colored bands denote 68\% CL intervals (from one or more observables), while the black lines denote $1, 2, (3) \sigma$ contours from the global fit.}.     \label{fig:BKnunuAligmentPlot} 
\end{figure}

\subsubsection*{Fit to current data}

The interplay of   $\cB(B^+ \rightarrow K^+ \nu\bar\nu)$, $\cB(K^+ \rightarrow \pi^+ \nu\bar\nu)$, collider and 
EW observables is controlled by the values of 
the two semi-leptonic WCs $\cC_{\ell q}^{(1)[3333]}$ and $\cC_{\ell q}^{(3)[3333]}$ and by the flavor-alignment parameter $\epsf$, which enters via the mixing matrix $L_d$  in Eq.~(\ref{eq:Ld}).
The $B^+ \rightarrow K^+ \nu\bar\nu$ and $K^+ \rightarrow \pi^+ \nu\bar\nu$ amplitudes scale with different powers of $\epsf$ (linear for $B\rightarrow K$ and quadratic for $K\rightarrow \pi$), while EWPO and collider observables are insensitive to $\epsf$.  
The most relevant EWPO constraining  $\cC_{\ell q}^{(1)[3333]}$ and $\cC_{\ell q}^{(3)[3333]}$ are 
$R_\tau$ and LFU tests in $\tau$ decays, while the most constraining collider observable is $\sigma(pp\to \tau \bar\tau)$.

In the upper-left panel of 
\cref{fig:BKnunuAligmentPlot}  we show the constraints from all these observables using current data in the  $\epsf$-- $\cC_{\ell q}^{(3)[3333]}$ plane, setting $\cC_{\ell q}^{(1)[3333]}=0$.
As can be seen, the compatibility of rare modes and EWPO requires $\epsf \gtrsim 0.5$ and improves as $\epsf$ increases.
Setting for simplicity $\epsf = 1$ and performing a 2D fit to the data with the two WCs as free parameters leads to the result in the upper-right panel of \cref{fig:BKnunuAligmentPlot}. With current data, the fit indicates a mild $2\sigma$ tension with respect to the SM. The best-fit point is $(\cC_{\ell q}^{(1)[3333]},\cC_{\ell q}^{(3)[3333]}) = (0.6,2.3) \times 10^{-7}~{\rm GeV}^{-2}$, corresponding to an effective NP scale of roughly 2 TeV.  

It is important to stress that collider and flavor observables each probe a single combination of $\cC_{\ell q}^{(1)[3333]}$ and $\cC_{\ell q}^{(3)[3333]}$, while EWPO are sensitive to each of them individually. Indeed, EWPO play a crucial role in selecting one of the two potential solutions allowed by collider and flavor data. This interplay among observables from different sectors is therefore an essential tool to determine the electroweak structure of the hypothetical NP sector. 
 Similarly, the different sensitivity of $B^+ \rightarrow K^+ \nu\bar\nu$ and $K^+ \rightarrow \pi^+ \nu\bar\nu$ to the alignment parameter $\epsf$ is essential to determine the flavor structure of NP. 

\subsubsection*{Future flavor and collider projections}

As previously mentioned, significant progress is expected in both $B^+ \rightarrow K^+ \nu\bar\nu$ and $K^+ \rightarrow \pi^+ \nu\bar\nu$ in the next 10-15 years. The Belle II experiment aims to probe $\mathcal{B}(B^+ \rightarrow K^+ \nu\bar\nu)$ with a 10\% relative error~\cite{Belle-II:2018jsg}, while NA62/HIKE will provide a final measurement of $\mathcal{B}(K^+ \rightarrow \pi^+ \nu\bar\nu)$ at the 5\% level~\cite{HIKE:2022qra}. Meanwhile, the high-luminosity phase of the LHC is currently on track to reach 3~${\rm ab}^{-1}$ of integrated luminosity, marking a 20-fold increase in available data. To illustrate the potential of these forthcoming improvements in flavor and collider observables, we perform a future projection of the results presented in the previous section. To do this, we fix the experimental measurements to the values suggested by the NP best-fit point determined with current data and then rescale the errors according to the final projected experimental performance. The outcome of this projection is shown in the bottom-right panel of \cref{fig:BKnunuAligmentPlot}. The results indicate that the combination of future flavor and collider observables, together with current EW data, would allow us to exclude the SM at more than $5\sigma$. Using the same projection methodology, we also display in the bottom-left panel of \cref{fig:BKnunuAligmentPlot} the projected preferred regions in the $\cC_{\ell q}^{(3)[3333]}$ vs. $\epsf$ plane, with $\cC_{\ell q}^{(1)[3333]}$ set to the best-fit point obtained with current data. This analysis suggests that future data will indeed allow us to probe the value of the alignment parameter, and potentially rule out values of $\epsf \lesssim 0.6$ at 95\% confidence level. An even more precise determination of $\epsf$ may become feasible with future improvements in EWPO.

   \begin{table}[p]
 \centering
\renewcommand{\arraystretch}{1.3} 
 \begin{tabular}{c|c|c|c}
Observable & Current Rel. Error ($10^{-3}$) & FCC-ee Rel. Error ($10^{-3}$) & Proj. Error Reduction \\
\hline
\hline 
$ \mathrm{  \Gamma_Z} $ &  2.3 & 0.1 & 23 \\
\hline
$ \sigma_{\rm had}^0$ &  37 & 5 & 7.4  \\
$ R_b $ &  3.06 & 0.3 & 10.2    \\
$ R_c $ &  17.4 & 1.5 & 11.6    \\
$ A_{\rm FB}^{0,b} $ &  15.5 & 1 & 15.5    \\
$ A_{\rm FB}^{0,c} $ &  47.5 & 3.08 & 15.4    \\
$ A_b $ &  21.4 & 3 & 7.13    \\
$ A_c $ &  40.4 & 8 & 5.05    \\
\hline
$ R_e $ &  2.41 & 0.3 & 8.03    \\
$ R_\mu $ &  1.59 & 0.05 & 31.8    \\
$ R_\tau $ &  2.17 & 0.1 & 21.7    \\
$ A_{\rm FB}^{0,e} $ &  154 & 5 & 30.8    \\
$ A_{\rm FB}^{0,\mu} $ &  80.1 & 3 & 26.7    \\
$ A_{\rm FB}^{0,\tau} $ &  104.8 & 5 & 21    \\
$ A_e^{**}$ &  14.3 & 0.11 & 130    \\
$ A_\mu^{**} $ &  102 & 0.15 & 680    \\
$ A_\tau^{**} $ &  102 & 0.3 & 340    \\
\hline
 \end{tabular}
 \caption{Projected FCC-ee improvement for $Z$-pole observables from~\cite{DeBlas:2019qco}. The $A_\ell^{**}$ are from lepton
polarization and LR asymmetry measurements at SLC.}
 \label{tab:FCCeePROJZpole}
 \end{table}

   \begin{table}[p]
 \centering
\renewcommand{\arraystretch}{1.3} 
 \begin{tabular}{c|c|c|c|c}
Observable & Value  & Error  & FCC-ee Tot. & Proj. Error Red. \\
\hline
\hline 
$ \mathrm{  \Gamma_W  ~(MeV) } $ &  2085 & 42 & 1.24 & 34 \\
$ \mathrm{  m_W  ~(MeV) } $ &  80350 & 15 & 0.39 & 38 \\
$ \tau \rightarrow \mu \nu\nu (\%)$ &  17.38 & 0.04 & 0.003 & 13 \\
$ {\rm Br}(W\rightarrow e \nu) (\%)$ & 10.71 & 0.16 & 0.0032 & 50 \\
$ {\rm Br}(W\rightarrow \mu \nu) (\%)$ & 10.63 & 0.15 & 0.0032 & 47 \\
$ {\rm Br}(W\rightarrow \tau \nu) (\%)$ & 11.38 & 0.21 & 0.0046 & 46 \\
$ \mu_{ b\bar b}$ &  0.99 & 0.12 & 0.003 & 40 \\
$ \mu_{c\bar c}$ &  8 & 22 & 0.022 & 1000 \\
$ \mu_{\tau\bar \tau} $ & 0.91 & 0.09 & 0.009 & 10 \\
$ \mu_{\mu \bar \mu} $ & 1.21 & 0.35 & 0.19 & 1.84 \\
\hline
 \end{tabular}
 \caption{Projected FCC-ee improvement for selected $H$, $\tau$ and $W$-pole observables from~\cite{DeBlas:2019qco,Blondel:2021ema,Bernardi:2022hny}.}
 \label{tab:FCCeePROJother}
 \end{table}

\subsection{Future circular $e^+ e^-$ colliders (FCC-ee)}

Having discussed the significance 
of flavor-physics measurements in the near future using a specific example, we now proceed with a systematic analysis of the  potential of a future circular $e^+ e^-$ collider for EW observables. 
Our focus is on the FCC-ee program, but our findings are generally applicable to other tera-$Z$ machines, such as the proposed CEPC project. We begin by compiling the projected FCC-ee improvements for $Z$-pole observables  in~\cref{tab:FCCeePROJZpole}~\cite{DeBlas:2019qco}. Projected improvements in $W$-pole, Higgs, and $\tau$-decay observables are shown in~\cref{tab:FCCeePROJother}~\cite{DeBlas:2019qco,Blondel:2021ema,Bernardi:2022hny}. A tera-$Z$ machine would provide $O(10^5)$ more $Z$-bosons than LEP, hence, in principle, statistical uncertainties on EWPO could improve by up to a factor of 300 (even more for observables that were not accessible at LEP, such as Higgs signal strengths).
  In practice, due to systematic and theoretical uncertainties, a factor of 10-100 (10) is typically expected for leptonic (hadronic) observables. With these promising improvements at hand, we construct a projected EW likelihood for FCC-ee by setting the experimental value of each EW observable to its corresponding SM theory prediction\footnote{For the SM theory predictions, we use the values given in \cite{Breso-Pla:2021qoe}.} and assuming the relative error is reduced as in the ``Proj. Error Reduction" column of~\cref{tab:FCCeePROJZpole,tab:FCCeePROJother}.

\subsubsection*{FCC-ee reach to probe the dominantly third-generation NP hypothesis}
\begin{figure}[t]
    \centering
    \includegraphics[width=\textwidth]{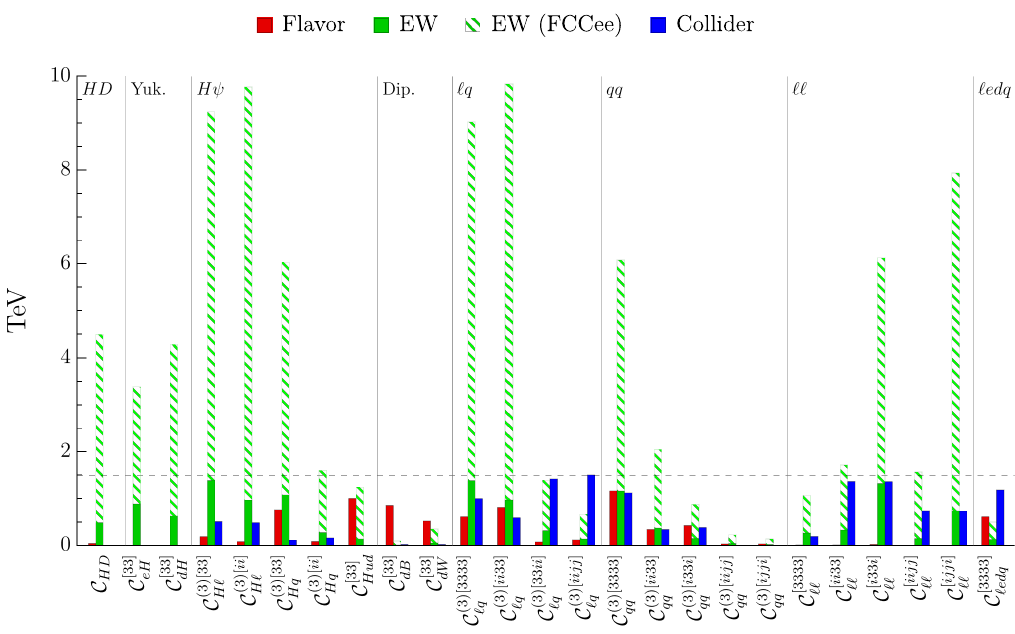}
    \caption{Bounds on the same operators shown in Fig.~\ref{fig:plot4}, including projections for FCC-ee.}
    \label{fig:plot4FCC}
\end{figure}

We first apply our FCC-ee likelihood to derive bounds on the selected operators discussed in ~\cref{sect:third}. These constraints are obtained after applying the $\varepsilon_i$ factors (see~\cref{fig:plot4}), 
i.e.~in the scenario with NP around the TeV scale coupled 
dominantly to the third generation. The results are shown in~\cref{fig:plot4FCC}. This analysis reveals that FCC-ee holds the potential to significantly improve the bounds on a large fraction of the representative set of effective operators. The strongest increase in sensitivity occurs for purely third-family four-fermion operators, such as $\cC_{\ell q}^{(3)[3333]}$ and $\cC_{qq}^{(3)[3333]}$,  whose current bounds are suppressed only by $\epsf$. 
These operators are strongly constrained by FCC-ee due to their $y_t$-enhanced running into $Z$-pole observables.
Additionally, the leptonic Higgs bi-fermion operators $\cC_{H\ell}$ are also strongly constrained, despite the $\varepsilon_H$ and $\varepsilon_L$ suppression factors. This is due to the expected improvement factor of approximately $O(100)$ in the $A_\ell$ EWPO at FCC-ee (see \cref{tab:FCCeePROJZpole}). 
While current data are compatible with TeV-scale NP within this setup,  
leaving open the possibility of a quasi-natural solution to the EW hierarchy 
problem, our projections show that FCC-ee will be able to 
completely rule out this possibility or, from a more optimistic perspective, 
provide clear evidence for this NP framework.

\subsubsection*{Future projections for operators most strongly probed by EWPO}
\begin{figure}
    \centering
    \includegraphics[width=\textwidth]{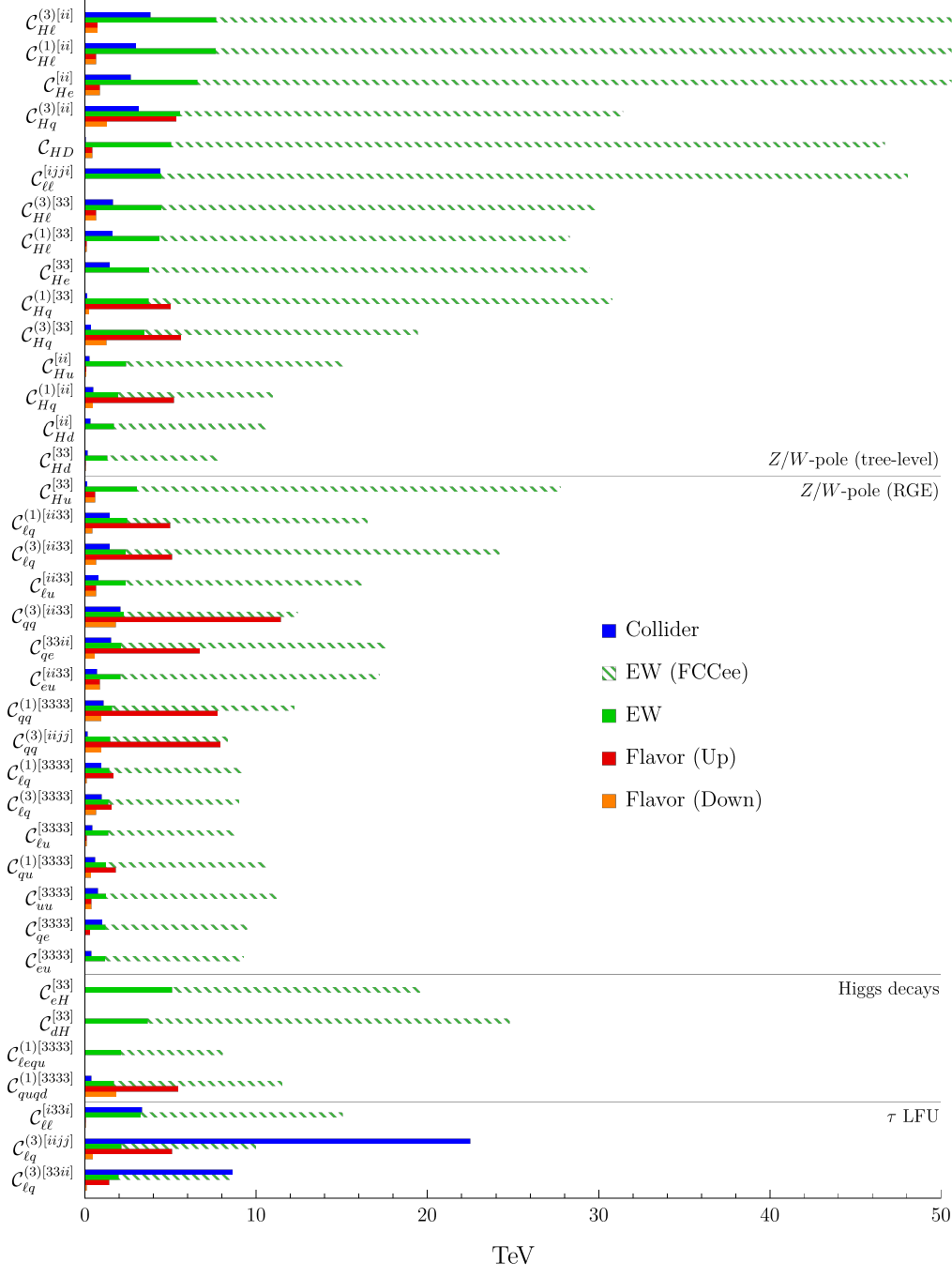}
    \caption{ FCC-ee projection for operators with the strongest current EW constraints, sorted by observable categories. The $Z/W$-pole (RGE) category contains operators which enter $Z/W$-pole observables only via RG mixing into $Z/W$-pole (tree-level) operators.} 
    \label{fig:EWplotvertical}
\end{figure}

We now proceed to a more  general  analysis, moving away from the 
third-generation NP scenario. We sort all operators based on the strength of current EW constraints and perform the FCC-ee projection, showing the results in~\cref{fig:EWplotvertical}. 

As expected, operators entering $Z$-pole observables at tree-level -- those involving Higgses -- receive extremely stringent bounds. In most instances, these bounds exceed $30$ TeV and, in some cases, even reach $50$ TeV. Moreover, many non-Higgs operators involving third-family quarks still receive bounds of 
$O(10)$~TeV due to their running into $Z$-pole observables, as shown in the $Z$-pole (RGE) category of~\cref{fig:EWplotvertical}. 

Two important comments are in order concerning these results. The first, which calls back to the pre-LHC message advocated in Ref.~\cite{Barbieri:2000gf}, is that a future EW precision machine such as FCC-ee is the best way to probe NP with sizable couplings to the Higgs, as always occurs in low-scale flavor models and/or models that solve the EW hierarchy problem. With FCC-ee, we can (indirectly) probe these well-motivated 
extensions of the SM up to effective scales of around $50$ TeV. This offers the possibility of discovering NP responsible for stabilizing the Higgs mass or, conversely, would put severe pressure on the concept of naturalness. Secondly, even NP that does not couple directly to the Higgs but does couple to the third generation will be probed up to effective scales of around $10$ TeV. As we mentioned, this is due to the strong running of top quark operators into $Z$-pole observables. 
In summary, the overall takeaway message is that FCC-ee has the potential to push 
most of the existing bounds on NP from the EW sector by one order of magnitude, positioning FCC-ee as a powerful indirect discovery machine for a broad class of NP scenarios.

%-----------------------------------------------------------------------------
\section{Conclusions} \label{sec:conclusions}
%-----------------------------------------------------------------------------

The SMEFT provides a  powerful   framework to describe high-scale extensions of the SM in general terms. However, to investigate more specific questions about the nature of NP, 
such as its flavor and electroweak structure, or the lowest energy scale at which it might manifest, 
the SMEFT alone is not sufficient. In order to address these questions, we 
must supplement the SMEFT with additional hypotheses.

In this paper, we have presented the first comprehensive analysis of 
experimental constraints on SMEFT operators under the general hypothesis that
NP couples universally to the two lightest fermion families, corresponding to a $U(2)^5$ symmetry acting upon them. 
We scrutinize electroweak, flavor, and collider data to derive   bounds on all 124 effective operators associated with the dimension-6 SMEFT basis in the $U(2)^5$ invariant limit. These limits are reported in Tables~\ref{tab:psi2H2D}-\ref{tab:bosonic} as conservative lower bounds on the effective NP scale pertaining to each operator. 
A distinctive aspect of our study is the consistent inclusion of RG effects, 
allowing us to establish connections between observables characterized by significantly different energy scales 
and express them all in terms of UV conditions on the SMEFT WCs. 

The primary physics motivation behind this study is to answer the question of whether present data are compatible with the hypothesis of new dynamics primarily coupled to the  third generation of SM fermions 
around the TeV scale and, if so,
under which UV conditions.  The answer to this question is illustrated through the 
sequence of plots in Figure~\ref{fig:plot1}--\ref{fig:plot4}.
Employing only the hypothesis of exact $U(2)^5$ symmetry, 
many operator bounds still exceed several~TeV.
However, with additional simple assumptions, such as the mild suppression
of all the operators involving light fields at the high scale,  
all present bounds can be reconciled with an effective scale  as low as 1.5~TeV. 
We stress that the  single-operator bounds,
 and the determination of the $\varepsilon_i$ suppression 
factors,
derived in this numerical 
analysis should be viewed with some degree of caution; they primarily serve to demonstrate the existence of a well-motivated class of NP models at nearby energy scales.
This result is  particularly encouraging in light of ongoing experimental searches for NP,
both at the low- and high-energy frontier. 

Our second key objective is to estimate the progress we can anticipate in the exploration of this motivated  NP setup in the medium and long term. To this end, we take into account the expected progress on flavor and collider observables in the near future, as well as the potential contribution of a  future circular $e^+ e^-$ collider. 
As exemplified by our discussion of the rare 
$B \to K \nu\bar\nu$ and $K\to \pi \nu\bar\nu$ decays, illustrated in Figure~\ref{fig:BKnunuAligmentPlot}, 
flavor observables remain a promising avenue for NP discovery. 
The example we have presented is representative of the broader role of flavor-changing processes in probing this class of NP models: to determine the flavor orientation of the third-generation, a defining characteristic of these models, improved precision in flavor observables is essential.

As far as EW observables are concerned, we have shown the substantial impact that a future $e^+ e^-$ collider, 
such as FCC-ee, could have. This is unsurprising given that, once RG effects are taken into account, 
123 out of the 124 operators in our basis are constrained by  EWPO (although they may not 
necessarily provide the most stringent constraints at present).
Moreover, many EWPO results are still dominated by LEP data. 
As illustrated in Figure~\ref{fig:EWplotvertical}, 
FCC-ee has the potential to 
improve most existing bounds by an order of magnitude or even more.
This could either disprove or provide clear evidence for a wide class of compelling 
NP models that are fully compatible with present data. We hope that these results will serve to illustrate the power of a future $e^+ e^-$ 
circular collider as indirect discovery machine for a broad class of NP scenarios.

%-----------------------------------------------------------------------------
\section*{Acknowledgments}
%-----------------------------------------------------------------------------
We would like to thank Olcyr Sumensari for useful discussions.
This project has received funding from the European Research Council~(ERC) under the European Union's Horizon~2020 research and innovation programme under grant agreement 833280~(FLAY), and by the Swiss National Science Foundation~(SNF) under contract~200020\_204428. The research of CC was supported by the Cluster of Excellence \textit{Precision Physics, Fundamental Interactions, and Structure of Matter} (PRISMA$^+$, EXC 2118/1) within the German Excellence Strategy (Project-ID 39083149).

\appendix

\section{Theory expressions for observables}
\label{app:B}

Here we collect the expressions for the relevant observables used in our analysis in terms of SMEFT and/or LEFT Wilson coefficients.

\subsection{LFU tests in $\tau$ decays}
In terms of LEFT WCs, the LFU ratios defined in   \eqref{eq:TauLFU_Definition} are given by: 
\begin{align}
\bigg|\frac{g_\tau}{g_\mu} \bigg| & \approx 1 - \frac{v^2}{2} \, {\rm Re} \Big(   [L_{\nu e}^{V,LL}]^{\rm NP}_{3113} - [L_{\nu e}^{V,LL}]^{\rm NP}_{2112} \Big) \,, \label{eq:TauLFUThVal} \\
\bigg|\frac{g_\tau}{g_e} \bigg| & \approx 1 - \frac{v^2}{2} \, {\rm Re} \Big(   [L_{\nu e}^{V,LL}]^{\rm NP}_{3223} -  [L_{\nu e}^{V,LL}]^{\rm NP}_{2112} \Big) \,, \nonumber \\
\bigg|\frac{g_\tau}{g_\mu} \bigg|_\pi & \approx  1 - \frac{v^2}{2} {\rm Re} \left( \frac{1}{V_{ud}^*} \left[ [L_{\nu e d u}^{V,LL}]^{\rm NP}_{3311}-  [L_{\nu e d u}^{V,LL}]^{\rm NP}_{2211}+  \chi_{\pi}^\tau [L_{\nu e d u}^{S,RL}]^{\rm NP}_{3311} -\chi_{\pi}^\mu [L_{\nu e d u}^{S,RL}]^{\rm NP}_{2211}  \right] \right)  \,,\nonumber \\
\bigg|\frac{g_\tau}{g_\mu} \bigg|_K & \approx  1 - \frac{v^2}{2} {\rm Re} \left( \frac{1}{V_{us}^*} \left[   [L_{\nu e d u}^{V,LL}]^{\rm NP}_{3321}-  [L_{\nu e d u}^{V,LL}]^{\rm NP}_{2221} +   \chi_{K}^\tau [L_{\nu e d u}^{S,RL}]^{\rm NP}_{3321} -\chi_{K}^\mu [L_{\nu e d u}^{S,RL}]^{\rm NP}_{2221} \right] \right)\,, \nonumber
%%%
\end{align}
where $\chi_\pi^{\tau,\mu}$ and $\qquad \chi_K^{\tau,\mu}$ are the chiral enhancement factors 
\begin{equation}
\chi_\pi^{\tau,\mu}   =  \frac{m_\pi^2}{m_{\tau,\mu}(m_d + m_u)} \,, \qquad \qquad \chi_K^{\tau,\mu}   =  \frac{m_K^2}{m_{\tau,\mu}(m_s + m_u)}  \,.
\end{equation}
While the vector operators do not run in the LEFT, the scalar operators receive a multiplicative factor of $\eta_S = 1.57$ when running from the EW scale down to $m_\tau$. In the down-quark mass basis, the SMEFT matching conditions read
\begin{align}
[L_{\nu e}^{V,LL}]^{\rm NP}_{\alpha\beta\beta\alpha}(m_\tau) = & -2 \sum_{\gamma = \alpha,\beta} \cC_{H\ell}^{(3)[\gamma\gamma]}(m_Z) + \cC_{\ell\ell}^{[\alpha\beta\beta\alpha]} (m_Z)+\cC_{\ell\ell}^{[\beta\alpha\alpha\beta]}(m_Z) \,,
\label{eq:LnueMatching}
\end{align}
\begin{align}
	[L_{\nu edu}^{V,LL}(m_\tau)]_{\alpha\alpha 11} & = 2V_{ui}^*\left[ \cC_{\ell q}^{(3)[\alpha\alpha1i]} - \cC_{H q}^{(3)[1i]} - \cC_{H \ell}^{(3)[\alpha\alpha]}\delta_{1i}\right] \,,  \label{eq:TauLFUThMatchingLEFTtoSMEFT} \\
	[L_{\nu edu}^{V,LL}(m_\tau)]_{\alpha\alpha 21} & = 2V_{ui}^*\left[ \cC_{\ell q}^{(3)[\alpha\alpha2i]} - \cC_{H q}^{(3)[2i]} - \cC_{H \ell}^{(3)[\alpha\alpha]}\delta_{2i}\right] \,, \nonumber \\
	[L_{\nu edu}^{S,RL}(m_\tau)]_{\alpha\alpha 11} & = \eta_S V_{ui}^* \, \cC_{ledq}^{[\alpha\alpha 1 i]} \,, \nonumber \\
	[L_{\nu edu}^{S,RL}(m_\tau)]_{\alpha\alpha 21} & = \eta_S V_{ui}^* \, \cC_{ledq}^{[\alpha\alpha 2 i]} \,. \nonumber
\end{align}
For the experimental values, we use the HFLAV averages~\cite{HeavyFlavorAveragingGroup:2022wzx}
\begin{align}
\begin{aligned}
(g_\tau / g_\mu)_\tau &= 1.0009 \pm 0.0014 \,, \\
(g_\tau / g_e)_\tau &= 1.0027 \pm 0.0014 \,, \\
%(g_\mu / g_e)_\tau &= 1.0019 \pm 0.0014 \,, \\
(g_\tau / g_\mu)_\pi &= 0.9959 \pm 0.0038 \,, \\
(g_\tau / g_\mu)_K &= 0.9855 \pm 0.0075 \,,
\label{eq:TauLFUExpVal}
\end{aligned}
\end{align}
with the following correlation matrix, ordered according 
the list of data in Eq.~(\ref{eq:TauLFUExpVal}):
\begin{equation}
\rho_{\tau} = \left(
\begin{array}{cccc}
 1 & 0.51  & 0.16 & 0.12 \\
 0.51 & 1 &  0.18 & 0.11 \\
 0.16 & 0.18  & 1 & 0.07 \\
 0.12 & 0.11  & 0.07 & 1 \\
\end{array}
\right) \,.
\label{eq:TauLFUExpCorr}
\end{equation}

\subsection{$H \to f\bar f$}

The effective Lagrangian for Higgs decays in the mass basis can be written as
\begin{align}
    \mathcal{L} = -\frac{1}{\sqrt{2}} h \sum_{f = \ell, u, d} y_{\rm eff}^{f,\,ij} \bar f_L^i f_R^j + {\rm h.c.} \,,
\end{align}
where we sum over flavor indices, and
\begin{align}
\begin{aligned}
    y_{\rm eff}^{\ell,\,ij} &= \frac{\sqrt{2} m^\ell_i}{v} \delta_{ij} - v^2 [\mathcal{C}_{eH}]_{ij} \,, \\
    y_{\rm eff}^{u,\,ij} &= \frac{\sqrt{2} m^u_i}{v} \delta_{ij} - v^2 [\mathcal{C}_{uH}]_{ij} \,, \\
    y_{\rm eff}^{d,\,ij} &= \frac{\sqrt{2} m^d_i}{v} \delta_{ij} - v^2 [\mathcal{C}_{dH}]_{ij} \,.
\end{aligned}
\end{align}
The signal strengths for $H\to f\bar f$ can then be expressed as
\begin{align}
    \mu_{ii} = \frac{\mathcal{B}(h\to f_i \bar f_i)}{\mathcal{B}(h\to f_i \bar f_i)_{\rm SM}} = \left| 1 - \frac{v^3}{\sqrt{2}m^f_i} [\mathcal{C}_{fH}]_{ii} \right|^2\,.
\label{eq:HiggsThVal}
\end{align}
The experimental inputs are summarized in Table \ref{tab:HiggsExpVal}.
\begin{table}
    \centering
    \begin{tabular}{|c|c|}
        \hline
        $bb$ & 0.99 $\pm$ 0.12 \\
        $cc$ & 8 $\pm$ 22 \\
        $\tau\tau$ & 0.91 $\pm$ 0.09 \\
        $\mu\mu$ & 1.21 $\pm$ 0.35 \\
        \hline
    \end{tabular}
    \caption{Experimental values for the Higgs signal strengths (from PDG~\cite{ParticleDataGroup:2022pth}).}
    \label{tab:HiggsExpVal}
\end{table} 

\subsection{$B \rightarrow X_s \gamma$}
\label{app:bsgammaNP}

Following Ref.~\cite{Misiak:2020vlo}, and defining the following effective Lagrangian
\begin{equation}
\mathcal{L}_{\rm eff} = -\frac{4G_F}{\sqrt{2}} V_{tb} V_{ts}^* \frac{m_b}{16\pi^2} \Big[ e\, C_{7\gamma s} (  \bar s_L \sigma^{\mu\nu} b_R) F_{\mu\nu} + g_s\, C_{8g s} (\bar s_L \sigma^{\mu\nu} T^a b_R) G_{\mu\nu}^a \Big] \,,
\end{equation}
the theory expression for 
$\mathcal{B}_{s\gamma} \equiv \cB(B \to X_{s}\gamma)$ reads
\begin{align}
10^4 \mathcal{B}_{s\gamma}  & = (3.39 \pm 0.17) - 2.10\, {\rm Re}\left(3.93 \, C_{7\gamma s} + C_{8g s}\right)_{\mu_{\rm EW}} 
 \,,
\end{align}
where we set $\mu_{\rm EW} = 160$ GeV.
For the experimental value, we use the current world average~\cite{Misiak:2020vlo,HFLAV:2019otj}
\begin{align} \label{bsg}
\mathcal{B}_{s\gamma}^{\rm exp} =& (3.32 \pm 0.15)\times 10^{-4}\,.
\end{align}
The matching of $C_{7\gamma s}$ and $C_{8g s}$ to the corresponding LEFT Wilson coefficients is
\begin{align}
C_{7\gamma s} = \frac{1}{e}\frac{8\pi^2 v^2}{m_b V_{tb} V_{ts}^*} [L_{d\gamma}]_{sb} \,, \hspace{15mm} C_{8g s} = \frac{1}{g_s}\frac{8\pi^2 v^2}{m_b V_{tb} V_{ts}^*} [L_{dG}]_{sb} \,,
\end{align}
where $[L_{d\gamma}]_{sb}$ and $[L_{dG}]_{sb}$ are understood to be in the mass eigenbasis. We compute them using the 1-loop SMEFT-LEFT matching conditions provided as a \texttt{Mathematica} notebook by the authors of Ref.~\cite{Dekens:2019ept}.

\subsection{$b\rightarrow c (u) \tau \nu $}
\label{app:FCCC}
The effective Lagrangian relevant for $b\rightarrow c \tau \nu$ transitions at the scale $\mu = m_b$ is
\begin{equation}
	\mathcal{L}_{\rm eff} \supset -\frac{2}{v^2}  V_{cb} \sum_i \cC_i \cO_i~,
	\label{eq:LeffDF1}
\end{equation}
where the operators are defined as
\begin{align}
&\cO_{V_L}  =(\overline{c}\gamma^{\mu}P_{L}b)(\overline{\tau}\gamma_{\mu}P_{L}\nu)\, , & \cO_{V_R}=  (\overline{c}\gamma^{\mu}P_{R}b)(\overline{\tau}\gamma_{\mu}P_{L}\nu)\, , \nonumber\\
&\cO_{S_L}  =(\overline{c}P_{L}b)(\overline{\tau}P_{L}\nu)\, , & \cO_{S_R}=  (\overline{c}P_{R}b)(\overline{\tau}P_{L}\nu)\, ,\\
& \cO_{T}  =(\overline{c}\sigma^{\mu\nu}P_{L}b)(\overline{\tau}\sigma_{\mu\nu}P_{L}\nu)\, . \nonumber
\label{eq:Opbctaunu}
\end{align}
The corresponding Wilson coefficients can be written in terms of the LEFT ones as~\cite{Gherardi:2020qhc}
\begin{align}
	\cC_{V_L} & = - \frac{v^2}{2 V_{cb}}[L_{\nu edu}^{V,LL}(m_b)]_{\nu_\tau \tau bc}^{*}\, , &
	\cC_{V_R} & = - \frac{v^2}{2 V_{cb}}[L_{\nu edu}^{V,LR}(m_b)]_{\nu_\tau \tau bc}^{*}\, , \nonumber \\
	\cC_{S_L} & = - \frac{v^2}{2 V_{cb}}[L_{\nu edu}^{S,RR}(m_b)]_{\nu_\tau \tau bc}^{*}\, , & 
	\cC_{S_R} & = - \frac{v^2}{2 V_{cb}}[L_{\nu edu}^{S,RL}(m_b)]_{\nu_\tau \tau bc}^{*}\, ,\\
	\cC_{T} & = - \frac{v^2}{2 V_{cb}}[L_{\nu edu}^{T,RR}(m_b)]_{\nu_\tau \tau bc}^{*}\, . \nonumber
\end{align}
The expressions for the 
$R_D$ and $R_{D^*}$ ratios testing LFUV in terms of these 
coefficients can be found in~\cite{Iguro:2018vqb,Becirevic:2022bev}. 
For the theory predictions we use the FLAG average~\cite{FlavourLatticeAveragingGroupFLAG:2021npn}  and the HPQCD results~\cite{Harrison:2021tol}
for the $B\to D$ and $B\to D^*$ form factors, respectively. 
We also include the  branching fractions $\mathcal{B}(B_{u,c} \rightarrow \tau \nu)$, which are also  sensitive  to LFU contributions~\cite{Iguro:2018vqb,Fuentes-Martin:2019mun}
\begin{equation}
 \frac{\mathcal{B}(B_{c} \rightarrow \tau \nu)}{\mathcal{B}(B_{c} \rightarrow \tau \nu)_{\rm SM}} =
  ~|1+\cC_{V_L}-\cC_{V_R} + 4.33 (\cC_{S_R}-\cC_{S_L})|^2 \,,
 \end{equation}
 \begin{equation}
\label{eq:BsubC}
 \frac{\mathcal{B}(B_{u} \rightarrow \tau \nu)}{\mathcal{B}(B_{u} \rightarrow \tau \nu)_{\rm SM}} =
  ~|1+\cC_{V_L}^u-\cC_{V_R}^u + 3.75 (\cC_{S_R}^u-\cC_{S_L}^u)|^2 \,,
 \end{equation}
 where $\cC^u_X = \cC_X (c\rightarrow u)$.

We are interested in the matching to the SMEFT at the EW scale. The running of the vector operators from $m_b$ to $m_Z$ is at the sub-percent level and can be safely neglected. For the LEFT running of the scalar and tensor operators, we use
\begin{equation}
[L_{\nu edu}^{S,RL}(m_b)]_{\nu_\tau \tau bc} = 1.33 \, [L_{\nu edu}^{S,RL}(m_Z)]_{\nu_\tau \tau bc} \,,
\end{equation}
and
\begin{equation}
\begin{pmatrix}
[L_{\nu edu}^{S,RR}(m_b)]_{\nu_\tau \tau bc}  \\
[L_{\nu edu}^{T,RR}(m_b)]_{\nu_\tau \tau bc}
\end{pmatrix} = 
\begin{pmatrix}
1.33 & - 0.015 \\
0 & 0.915 
\end{pmatrix}
\begin{pmatrix}
[L_{\nu edu}^{S,RR}(m_Z)]_{\nu_\tau \tau bc}  \\
[L_{\nu edu}^{T,RR}(m_Z)]_{\nu_\tau \tau bc} 
\end{pmatrix}\,.
\end{equation}
At the EW scale, the matching conditions to the SMEFT are:
\begin{align}
\begin{aligned}
	[L_{\nu edu}^{V,LL}(m_Z)]_{\nu_\tau \tau bc} & = 2[L_d^\dagger]_{bi}\left[ \cC_{\ell q}^{(3)[33 ij]} - \cC_{H q}^{(3)[ij]} - \cC_{H \ell}^{(3)[33]}\delta_{ij}\right] [L_u]_{jc}\,,  \\
	[L_{\nu edu}^{V,LR}(m_Z)]_{\nu_\tau \tau bc} & = -[\cC_{Hud}^{[cb]}]^* \,,  \\
	[L_{\nu edu}^{S,RR}(m_Z)]_{\nu_\tau \tau bc} & = [L_d^\dagger]_{bi} \big( \cC_{lequ}^{(1) [33i2]} \big)  \,,  \\
	[L_{\nu edu}^{S,RL}(m_Z)]_{\nu_\tau \tau bc} & = \big( \cC_{ledq}^{[333i]} \big) [L_u]_{ic}  \,,  \\
	[L_{\nu edu}^{T,RR}(m_Z)]_{\nu_\tau \tau bc} & = [L_d^\dagger]_{bi} \big( \cC_{lequ}^{(3) [33i2]} \big)  \,.
 \end{aligned}
\end{align}
Note that $\cC_{Hud}$ and $\cC_{Hq}$ give LFU contributions, so they should be included in the matching for $\mathcal{B}(B_{u,c} \rightarrow \tau \nu)$, but not in the LFUV $R$-ratios. 

For the experimental data on $R_{D^{(*)}}$, we use the latest HFLAV averages~\cite{HFLAV:2019otj} 
\begin{align}
\begin{aligned}
R_{D^*}^{\rm exp} &= 0.284 \pm 0.012_{\rm total}\,, \\
R_{D}^{\rm exp} &= 0.357 \pm 0.029_{\rm total} \,, 
\end{aligned}
\end{align}
with correlation $\rho = -0.37$. Also for the SM predictions we use the HFLAV averages~\cite{HFLAV:2019otj}:
\begin{align}
    R_{D}^{\text{SM}} = 0.298(4)\,, \hspace{10mm}
    R_{D^*}^{\text{SM}} = 0.254(5)\,.
\end{align}
For the $B_u$ branching fraction, the data we use are 
$\mathcal{B}\left(B_u \rightarrow \tau \nu \right)_{\mathrm{exp}}=1.09(24) \times 10^{-4}$~\cite{ParticleDataGroup:2022pth}
 and $\mathcal{B}\left(B_u \rightarrow \tau \nu \right)_{\mathrm{SM}}=0.812(54) \times 10^{-4}$~\cite{Bona:2022zhn}. For the $B_c$ branching fraction, we use the upper bound $\mathcal{B}\left(B_c \rightarrow \tau \nu \right)_{\mathrm{exp}} < 0.1~(95\%~{\rm CL})$~\cite{Alonso:2016oyd,Akeroyd:2017mhr} and $\mathcal{B}\left(B_c \rightarrow \tau \nu \right)_{\mathrm{SM}}=0.023$ (theory uncertainty neglected).

\subsection{$b\rightarrow s \ell\ell $}
\label{app:bsllNP}
The low-energy effective Lagrangian describing $b\to s (d)  \ell \ell$ transitions is given by
\begin{align}
\label{eq:LWET}
\mathcal{L}_{b \to d_i \ell  \ell} & =  \frac{4G_F}{\sqrt2}\,V_{ti}^* V_{tb} \sum_{k=9,10} \, \mathcal{C}_k^{i \ell}\,\mathcal{O}_k^{i \ell} \,,
\end{align}
where
\begin{align}
\begin{aligned}
\label{eq:LWEToperators}
\mathcal{O}_9^{i \ell}&= \frac{\alpha}{4 \pi} \left(\bar d_L^i \gamma_\mu b_L \right) \left(\bar{\ell} \gamma^\mu \ell \right)\,,  \\[5pt] 
\mathcal{O}_{10}^{i \ell}&=\frac{\alpha}{4 \pi} \left(\bar d_L^i \gamma_\mu b_L \right)
\left(\bar{\ell} \gamma^\mu\gamma_5\, \ell \right)\,. 
\end{aligned}
\end{align}
These Wilson coefficients are to be interpreted as pure NP contributions. With this in mind, the matching of $\cC_9$ and $\cC_{10}$ to LEFT Wilson coefficients reads~\cite{Gherardi:2020qhc}
\begin{align}
\cC_{9(10)}^{ i\ell} &= \frac{\pi v^2}{\alpha V_{tb}V_{t i}^*} \bigg( [L_{de}^{V,LR}]_{i b\ell\ell} \pm [L_{ed}^{V,LL}]_{\ell\ell i b} \bigg) \,.
\end{align}
The tree-level expression in terms of SMEFT Wilson coefficients is then
\begin{align}
\begin{aligned}
\cC_9^{\alpha \ell} &= \frac{\pi v^2}{\alpha V_{tb}V_{ts}^*} [L_d^\dagger]_{\alpha i} \Big[ \cC_{qe}^{[ij\ell\ell]} + \cC_{\ell q}^{(1)[\ell\ell ij]} + \cC_{\ell q}^{(3)[\ell\ell ij]} - \zeta \left(\cC_{Hq}^{(1)[ij]} + \cC_{Hq}^{(3)[ij]} \right) \Big] [L_d]_{jb} \,, \\
\cC_{10}^{\alpha \ell} & = \frac{\pi v^2}{\alpha V_{tb}V_{ts}^*} [L_d^\dagger]_{\alpha i} \Big[ \cC_{qe}^{[ij\ell\ell]} - \cC_{\ell q}^{(1)[\ell\ell ij]} - \cC_{\ell q}^{(3)[\ell\ell ij]} + \cC_{Hq}^{(1)[ij]} + \cC_{Hq}^{(3)[ij]} \Big] [L_d]_{jb} \,,
\end{aligned}
\end{align}
where $\zeta = 1-4s_W^2$. The theory expression for $\mathcal{B}(B_{s(d)} \to \ell^+ \ell^- )$ is given by
\begin{align}
\mathcal{B}(B_{s(d)} \to \ell^+ \ell^- ) 
&= \mathcal{B}(B_{s(d)} \to \ell^+ \ell^-)_{\mathrm{SM}}     \left| 1 + \frac{\cC_{10}^{s(d) \ell}}{\cC_{10,\mathrm{SM}}} \right|^2 \,,
\label{eq:BmumuThVal}
\end{align}
with $\cC_{10,\mathrm{SM}} = -4.188$.
For $\mathcal{B}(B_{s,d}\rightarrow \mu\mu)$, we use the updated experimental averages given in Appendix C of Ref.~\cite{Greljo:2022jac}, while for the SM theory prediction we follow~\cite{Beneke:2019slt}. For $\cC_9$, we fit to the value $[C_9^{\rm U}]_{\rm exp} = -0.77 \pm 0.21$ given in Appendix B of Ref.~\cite{Greljo:2022jac} as a pseudo-observable proxy that averages various $b\to s\ell\ell$ measurements.

\subsection{$b(s)\rightarrow s(d) \nu\bar \nu$}
\label{app:bsnunuNP}

We define the low-energy effective Langrangian describing $d_{\alpha}\rightarrow d_{\beta} \nu\bar \nu$ transitions as
\begin{equation}
\mathcal{L}_{b\rightarrow s \nu\bar \nu} = \frac{4 G_F}{\sqrt{2}} V_{t\alpha}^* V_{t\beta} \frac{\alpha_W}{2\pi} C_L^{ij\ell} (\bar d_L^{\alpha} \gamma_{\mu} d_L^{\beta})(\bar \nu_L^\ell \gamma^\mu \nu_L^\ell) \,,
       \end{equation}
where $4\pi \alpha_W = g_L^2$.
Following \cite{Gherardi:2020qhc} and defining $C_L^{\alpha\beta\ell} \equiv C_{L, \rm SM}^{\alpha\beta\ell} + \Delta C_L^{\alpha\beta\ell}$, the NP contribution to $C_L^{\alpha\beta\ell}$ can be written in terms of the LEFT Wilson coefficients as
\begin{align}
    \Delta C_{L}^{\alpha\beta \ell} =  \frac{\pi v^2}{\alpha_W V_{t\alpha}^* V_{t\beta}} [L_{\nu d}^{V,LL}]_{\ell \ell \alpha\beta} \,.
\end{align}
In terms of the SMEFT WC's, we find:
\begin{align}
    [L_{\nu d}^{V,LL}]_{\ell \ell \alpha\beta} = [L_d^\dagger]_{\alpha i} \Big[ \cC_{\ell q}^{(1)[\ell\ell ij]} - \cC_{\ell q}^{(3)[\ell\ell ij]} + \cC_{Hq}^{(1)[ij]} + \cC_{Hq}^{(3)[ij]} \Big] [L_d]_{j\beta}   \,.
\end{align}

\paragraph*{I.}
For $b\rightarrow s \nu\bar\nu $ transitions we have 
\begin{align}
    \frac{\cB(B \to K^{(*)} \nu \nu)}{\cB(B \to K^{(*)} \nu \nu)_{\rm SM} }  = \frac{1}{3|C_{L, \rm SM}^{sb\ell}|^2}\sum_{\ell = e,\mu,\tau}  \big|C_{L, \rm SM}^{sb\ell}+\Delta C_L^{sb\ell}  \big|^2  \,,
\end{align}
where the SM contribution is $C_{L, \rm SM}^{sb\ell} = - 1.469 \pm 0.017$. For the experimental value, we use the combination reported in~\cite{Glazov}
\begin{equation}
\frac{\cB(B \to K^{(*)} \nu \nu)}{\cB(B \to K^{(*)} \nu \nu)_{\rm SM} } = 2.8 \pm 0.8 \,.
\end{equation}

\paragraph*{II.}
For $s\rightarrow d \nu\bar\nu $ transitions, we use the formulae given in~\cite{Buras:2015qea}
\begin{align}
\begin{aligned}
\frac{\cB(K^+ \to \pi^+ \nu \nu)}{\cB(K^+ \to \pi^+ \nu \nu)_{\rm SM} } &= \frac{1}{3|\mathcal{A}^{ds}_{\rm SM}|^2} \sum_{\ell=e,\mu,\tau}  \big|\mathcal{A}^{ds}_{\rm SM} +  \Delta C_L^{ds\ell}\big|^2 \,, \\
% \\
\frac{\cB(K_L \to \pi^0 \nu \nu)}{\cB(K_L \to \pi^0 \nu \nu)_{\rm SM}} &=
 \frac{1}{3 {\rm Im}\big[ V_{td}^* V_{ts} C_{L,\rm SM}^{ds\ell} \big]^2}\sum_{\ell=e,\mu,\tau} 
{\rm Im}\big[V_{td}^* V_{ts} (C_{L,\rm SM}^{ds\ell} + \Delta C_L^{ds\ell})\big]^2~, 
\label{eq:BRKLpnn}
\end{aligned}
\end{align}
where
\begin{equation}
\mathcal{A}^{ds}_{\rm SM} =  C_{L,\rm SM}^{ds\ell}  + \frac{|V_{us}|^5 |V_{cs}|}{V_{td}^* V_{ts}} P_c  
\end{equation}
is the full SM amplitude including the short-distance piece $C_{L,\rm SM}^{ds\ell}= - 1.48\pm 0.01$~\cite{Buchalla:1998ba}, and the long-distance contribution due to charm and light-quark loops encoded by $P_c =0.404 \pm 0.024$ (averaged over the different lepton species). The SM theory predictions are~\cite{Buras:2015qea}
\begin{align}
\begin{aligned}
\cB(K^+ \to \pi^+ \nu \nu)_{\rm SM} & = (9.11 \pm 0.72)\times 10^{-11} \,, \\
\cB(K_L \to \pi^0 \nu \nu)_{\rm SM} & = (3.00 \pm 0.30)\times 10^{-11} \,,
\end{aligned}
\end{align}
while for the experimental measurements we use~\cite{NA62:2021zjw,KOTO:2020prk}
\begin{align}
\begin{aligned}
\cB(K^+ \to \pi^+ \nu \nu)_{\rm exp} & = (10.6 \pm 3.8)\times 10^{-11} \,, \\
\cB(K_L \to \pi^0 \nu \nu)_{\rm exp} & < 2.98 \times 10^{-9}~(68\%~\text{CL}) \,.
\end{aligned}
\end{align}

\begin{table}[t]
\centering
\renewcommand{\arraystretch}{1.3} 
\begin{tabular}{|c|c|c|}
\hline
coefficient &    exp. bound ([$\mathrm{GeV}^{-2}$])\\
\hline 
\hline
$|C_{B_s}^1|$ & $(0.0 \pm 9.7) \times 10^{-12}$ \\
\hline
$|C_{B_d}^1|$ &  $(0.0 \pm 4.8) \times 10^{-13}$ \\
\hline
$\mathrm{Re}C_K^1$& $(0.5 \pm 3.7) \times 10^{-13}$  \\
\hline
$\mathrm{Im}C_K^1$ & $(0.6 \pm 0.9) \times 10^{-15}$\\
\hline
$\mathrm{Re}C_D^1$&  $(0.3 \pm 1.4) \times 10^{-13}$\\
\hline
$\mathrm{Im}C_D^1$ & $(-0.3 \pm 4.7) \times 10^{-15}$ \\
\hline
\end{tabular}
\caption{Experimental constraints on  the 
$\Delta F=2$  effective coefficients (from \cite{utfit}, symmetrizing the intervals and rescaling the widths to go from $95\%$ CL to $1 \sigma$).}
\label{tab:DF2expbounds}
\end{table}

\subsection{$\Delta F = 2$}
\label{app:DF2}
The low-energy effective Lagrangian describing meson mixing is
\begin{equation}
\mathcal{L}_{\Delta F = 2} = -\cC_{B_s}^1 (\bar s_L \gamma_\mu b_L)^2 - \cC_{B_d}^1 (\bar d_L \gamma_\mu b_L)^2 - \cC_{K}^1 (\bar d_L \gamma_\mu s_L)^2 - \cC_{D}^1 (\bar u_L \gamma_\mu c_L)^2 \,. 
\end{equation}
The matching conditions are
\begin{align}
\begin{aligned}
\cC_{B_s}^1 &= - [L_{dd}^{V,LL}]_{sbsb} = - [L_d^\dagger]_{si} [L_d^\dagger]_{sk} \left( \cC_{qq}^{(1)[ijkl]} + \cC_{qq}^{(3)[ijkl]} \right) [L_d]_{jb} [L_d]_{lb} \,,\\
\cC_{B_d}^1 &= - [L_{dd}^{V,LL}]_{dbdb} = - [L_d^\dagger]_{di} [L_d^\dagger]_{dk} \left( \cC_{qq}^{(1)[ijkl]} + \cC_{qq}^{(3)[ijkl]} \right) [L_d]_{jb} [L_d]_{lb}\,,  \\
\cC_{K}^1 &= - [L_{dd}^{V,LL}]_{dsds} = - [L_d^\dagger]_{di} [L_d^\dagger]_{dk} \left( \cC_{qq}^{(1)[ijkl]} + \cC_{qq}^{(3)[ijkl]} \right) [L_d]_{js} [L_d]_{ls} \,, \\
\cC_{D}^1 &= - [L_{uu}^{V,LL}]_{ucuc} = - [L_u^\dagger]_{ui} [L_u^\dagger]_{uk} \left( \cC_{qq}^{(1)[ijkl]} + \cC_{qq}^{(3)[ijkl]} \right) [L_u]_{jc} [L_u]_{lc} \,.
\end{aligned}
\end{align}
As experimental bounds we use the constraints on the WCs given on 
page~18 in \cite{utfit}, summarized in Table \ref{tab:DF2expbounds}.

\section{Individual operator bounds by sector}
\label{app:C}

We report here the bounds on the WCs derived by separate sets of observables: 
    flavor observables in the up-aligned basis (Table~\ref{tab:up_basis}), 
    flavor observables in the down-aligned basis (Table~\ref{tab:down_basis}), 
    EW observables (Table~\ref{tab:EW}), 
    collider observbles (Table~\ref{tab:Coll}).
For each WC the bounds are reported for the the two possible signs, with the indication of the most constraining observable in each direction.
For the sake of simplicity, in each table only the  most stringent bounds are reported, as indicated in the captions.

\begin{table}[t]
 \centering
\small{
\renewcommand{\arraystretch}{1.25} 
 \begin{tabular}{c|c|c|c|c|c}
 coeff. &  $\Lambda_{-}~[\text{TeV}]$ & obs. & $\Lambda_{+}~[\text{TeV}]$ & obs. & $\Lambda~[\text{TeV}]$ \\
\hline
\hline 
$\mathcal{C}_{dB}^{[33]}$ & $-82.9$ & $B \to X_s \gamma$ & $74.8$ & $B \to X_s \gamma $ & $74.8$  \\
$\mathcal{C}_{dW}^{[33]}$ & $-53.0$ & $B \to X_s \gamma$ &  $58.8$ & $B \to X_s \gamma$ &  $53.0$  \\
$\mathcal{C}_{dG}^{[33]}$ & $-28.3$ & $B \to X_s \gamma$ & $25.5$ & $B \to X_s \gamma$ & $25.5$  \\
$\mathcal{C}_{qq}^{(3)[ii33]}$ & $-12.1$ & $|C_{Bs}|$ & $11.5$ & $|C_{Bs}|$  & $11.5$  \\
$\mathcal{C}_{qq}^{(1)[i33i]}$ &  $-12.3$  & $|C_{Bs}|$ &  $11.3$  & $|C_{Bs}|$ &  $11.3$ \\
$\mathcal{C}_{qq}^{(3)[i33i]}$ & $-12.4$ & $|C_{Bs}|$ & $11.2$ & $|C_{Bs}|$ & $11.2$ \\
$\mathcal{C}_{qq}^{(1)[ii33]}$ & $-12.5$ & $|C_{Bs}|$ & $11.2$ &$|C_{Bs}|$  & $11.2$ \\
$\mathcal{C}_{qq}^{(3)[3333]}$ & $-8.2$ &$|C_{Bs}|$ & $8.4$  & $|C_{Bs}|$ & $8.2$ \\
$\mathcal{C}_{qq}^{(1)[ijji]}$ & $-8.1$ & $|C_{Bs}|$ & $8.8$ &$|C_{Bs}|$  & $8.1$ \\
$\mathcal{C}_{qq}^{(1)[iijj]}$ & $-8.1$ & $|C_{Bs}|$ & $8.7$ & $|C_{Bs}|$ & $8.1$ \\
$\mathcal{C}_{qq}^{(3)[ijji]}$ & $-8.0$ & $|C_{Bs}|$ & $8.8$ & $|C_{Bs}|$ & $8.0$ \\
$\mathcal{C}_{qq}^{(3)[iijj]}$ & $-7.9$ & $|C_{Bs}|$ & $8.9$ & $|C_{Bs}|$ & $7.9$ \\
$\mathcal{C}_{qq}^{(1)[3333]}$ & $-7.8$ & $|C_{Bs}|$ & $9.0$ & $|C_{Bs}|$ &  $7.8$ \\
$\mathcal{C}_{qe}^{[iijj]}$ & $-10.2$ & $\cC_9^U$ & $6.8$ & $B_s \to \mu^+ \mu^-$ & $6.8$  \\
$\mathcal{C}_{qe}^{[33ii]}$ & $-6.7$ & $B_s \to \mu^+ \mu^-$ & $10.1$ & $\cC_9^U$ & $6.7$ \\
$\mathcal{C}_{Hq}^{(3)[33]}$ & $-9.8$ & $B_s \to \mu^+ \mu^-$ & $5.6$ & $B_s \to \mu^+ \mu^-$ & $5.6$  \\
$\mathcal{C}_{quqd}^{(1)[3333]}$ & $-6.1$ & $B \to X_s \gamma$ & $5.5$ & $B \to X_s \gamma$ &  $5.5$ \\
$\mathcal{C}_{Hq}^{(3)[ii]}$ & $-5.3$ & $B_s \to \mu^+ \mu^-$ & $9.3$ & $B_s \to \mu^+ \mu^-$ & $5.3$ \\
$\mathcal{C}_{Hq}^{(1)[ii]}$ & $-5.2$ & $B_s \to \mu^+ \mu^-$ & $9.0$ & $B_s \to \mu^+ \mu^-$ & $5.2$  \\
$\mathcal{C}_{Hq}^{(1)[33]}$ & $-9.0$ & $B_s \to \mu^+ \mu^-$ & $5.2$ & $B_s \to \mu^+ \mu^-$ & $5.2$ \\
$\mathcal{C}_{quqd}^{(8)[3333]}$ & $-5.6$  & $B \to X_s \gamma$ &  $5.1$ & $B \to X_s \gamma$ & $5.1$ \\
$\mathcal{C}_{\ell q}^{(3)[iijj]}$ & $23$ & $K \to \pi \nu \bar{\nu}$ & $5.1$ & $B_s \to \mu^+ \mu^-$ &  $5.1$ \\
$\mathcal{C}_{\ell q}^{(3)[ii33]}$ & $-5.1$ & $B_s \to \mu^+ \mu^-$ & $-25.5$ & $K \to \pi \nu \bar{\nu}$ &  $5.1$ \\
$\mathcal{C}_{\ell q}^{(1)[iijj]}$ & $19.3$ & $R_{D^\ast}$ & $5.0$ & $B_s \to \mu^+ \mu^-$ &  $5.0$ \\
$\mathcal{C}_{\ell q}^{(1)[ii33]}$ & $-5.0$ & $B_s \to \mu^+ \mu^-$ & $-19.1$ & $K_L \to \pi_0 \nu \bar{\nu}$ &  $5.0$  \\
$\mathcal{C}_{Hud}^{[33]}$ & $-3.2$ & $B \to X_s \gamma$ & $3.6$ & $B \to X_s \gamma$ & $3.2$ \\
$\mathcal{C}_{qu}^{(1)[ii33]}$ & $-1.8$ & $B_s \to \mu^+ \mu^-$ & $2.6$ & $B_s \to \mu^+ \mu^-$ & $1.8$ \\
$\mathcal{C}_{qu}^{(1) [3333]}$ & $-2.5$ & $B_s \to \mu^+ \mu^-$ &  $1.8$ & $B_s \to \mu^+ \mu^-$ & $1.8$ \\
$\mathcal{C}_{\ell q}^{(1)[3333]}$ & $-1.7$ & $K \to \pi \nu \bar{\nu}$ & $6.3$ & $K \to \pi \nu \bar{\nu}$ & $1.7$\\
$\mathcal{C}_{\ell q}^{(1)[33ii]}$ & $-3.6$ & $K \to \pi \nu \bar{\nu}$ & $1.6$ & $K \to \pi \nu \bar{\nu}$ & $1.6$  \\
$\mathcal{C}_{\ell q}^{(3)[3333]}$ & $-1.5$ & $K \to \pi \nu \bar{\nu}$ & $-2.0$ & $K \to \pi \nu \bar{\nu}$ & $-2.0$\\
$\mathcal{C}_{\ell q}^{(3)[33ii]}$ & $-1.4$ & $K \to \pi \nu \bar{\nu}$ & $-7.0$ & $K \to \pi \nu \bar{\nu}$ & $1.4$ \\
$\mathcal{C}_{G}$ & $-1.1$ & $B \to X_s \gamma$ & $1.3$ & $B \to X_s \gamma$ & $1.1$  \\
$\mathcal{C}_{HWB}$ & $-1.1$ & $B \to X_s \gamma$ & $1.0$ & $B \to X_s \gamma$ & $1.0$  \\
\hline
 \end{tabular}
  }
 \caption{Bounds from flavor observables in the 
 up-aligned basis. Only bounds exceeding $1$~TeV are reported.} 
 \label{tab:up_basis}
 \end{table}

\newpage 

  \begin{table}[t]
   \centering
\small{
\renewcommand{\arraystretch}{1.25} 
 \begin{tabular}{c|c|c|c|c|c}
 coeff. &  $\Lambda_{-}~[\text{TeV}]$ & obs. & $\Lambda_{+}~[\text{TeV}]$ & obs. & $\Lambda~[\text{TeV}]$ \\
\hline
\hline 
$\mathcal{C}_{dW}^{[33]}$ & $-17.4$ & $B \to X_s \gamma$ & $15.7$ & $B \to X_s \gamma$  & $15.7$ \\
$\mathcal{C}_{dB}^{[33]}$ & $-16.8$ & $B \to X_s \gamma$ & $15.2$ & $B \to X_s \gamma$  & $15.2$ \\
$\mathcal{C}_{dG}^{[33]}$ & $-4.4$ & $B \to X_s \gamma$ & $4.0$ & $B \to X_s \gamma$  & $4.0$ \\
$\mathcal{C}_{Hud}^{[33]}$ & $-3.2$ & $B \to X_s \gamma$ & $3.6$ & $B \to X_s \gamma$  & $3.2$ \\
$\mathcal{C}_{qq}^{(3)[i33i]}$ & $-4.6$ & $B_s \to \mu^+ \mu^-$ & $2.6$ & $B_s \to \mu^+ \mu^-$ & $2.6$ \\
$\mathcal{C}_{quqd}^{(1)[3333]}$ & $-1.8$ & $B \to X_s \gamma$ & $2.0$ & $B \to X_s \gamma$ & $1.8$ \\
$\mathcal{C}_{qq}^{(3)[ii33]}$ & $-1.8$ & $B_s \to \mu^+ \mu^-$ & $3.0$ & $B_s \to \mu^+ \mu^-$ & $1.8$ \\
$\mathcal{C}_{qq}^{(1)[i33i]}$ & $-1.4$ & $B_s \to \mu^+ \mu^-$ & $2.0$ & $B_s \to \mu^+ \mu^-$ &  $1.4$ \\
$\mathcal{C}_{qq}^{(1)[ii33]}$ & $-1.4$ & $\mathrm{Im} C_{D_1}$ & $1.3$ & $\mathrm{Im} C_{D_1}$ & $1.3$\\
$\mathcal{C}_{Hq}^{(3)[ii]}$ & $-2.2$ & $B_s \to \mu^+ \mu^-$ & $1.3$ & $B_s \to \mu^+ \mu^-$ & $1.3$  \\
$\mathcal{C}_{Hq}^{(3)[33]}$ & $-2.1$ & $B_s \to \mu^+ \mu^-$ & $1.2$ & $B_s \to \mu^+ \mu^-$ & $1.2$ \\
$\mathcal{C}_{G}$ & $-1.1$ & $B \to X_s \gamma$ &  $1.3$ & $B \to X_s \gamma$ & $1.1$ \\
$\mathcal{C}_{qq}^{(3)[ijji]}$ & $-1.2$ & $K \to \pi \nu \bar{\nu}$ & $1.1$ & $\mathrm{Im} C_{D}$ & $1.1$ \\
$\mathcal{C}_{qq}^{(1)[ijji]}$ & $-1.1$ & $\mathrm{Im} C_{D}$ & $1.1$ & $\mathrm{Im} C_{D}$ & $1.1$ \\
$\mathcal{C}_{HWB}$ & $-1.1$ & $B \to X_s \gamma$ & $1.0$ & $B \to X_s \gamma$ & $1.0$ \\
$\mathcal{C}_{quqd}^{(8)[3333]}$ &  $-1.0$ &  $B \to X_s \gamma$ & $1.1$& $B \to X_s \gamma$ & $1.0$ \\
$\mathcal{C}_{qq}^{(3)[iijj]}$ & $-1.0$ & $\mathrm{Im} C_{D}$ & $1.0$ & $\mathrm{Im} C_{D}$ & $1.0$ \\
$\mathcal{C}_{qq}^{(3)[3333]}$ & $-1.0$ & $\mathrm{Im} C_{D}$ & $1.0$ & $\mathrm{Im} C_{D}$ & $1.0$ \\
$\mathcal{C}_{qq}^{(1)[3333]}$ & $-1.0$ & $\mathrm{Im} C_{D}$ & $1.0$ & $\mathrm{Im} C_{D}$ & $1.0$ \\
\hline
 \end{tabular}
 }
 \caption{Bounds from flavor observables in the 
 down-aligned basis. Only bounds exceeding $1$~TeV are reported.}
 \label{tab:down_basis}

 \end{table}

 \newpage 

   \begin{table}[t]
    \centering
   \small{
\renewcommand{\arraystretch}{1.25} 
\begin{tabular}{c|c|c|c|c|c}
coeff. & $\Lambda_{-}\text{ [TeV]}$ & Obs. & $\Lambda_{+}\text{ [TeV]}$ & Obs. & $\Lambda\text{ [TeV]}$\\ \hline\hline
$\mathcal{C}_{HWB}$ & -9. & $A_b^{\rm FB}$ & 17.1 & $m_W$ & 9. \\
$\mathcal{C}_{H\ell }^{(3)[ii]}$ & -7.7 & $\sigma_{\rm had}$ & 20.4 & $m_W$ & 7.7 \\
$\mathcal{C}_{H\ell }^{(1)[ii]}$ & -7.6 & $\sigma_{\rm had}$ & 10. & $\sigma_{\rm had}$ & 7.6 \\
$\mathcal{C}_{He}^{[ii]}$ & -9.6 & $\sigma_{\rm had}$ & 6.6 & $\sigma_{\rm had}$ & 6.6 \\
$\mathcal{C}_{Hq}^{(3)[ii]}$ & -8.6 & $\Gamma_Z$ & 5.6 & $R_\tau$ & 5.6 \\
$\mathcal{C}_{HD}$ & -5.1 & $A_b^{\rm FB}$ & 11.1 & $m_W$ & 5.1 \\
$\mathcal{C}_{eH}^{[33]}$ & -7.8 & $H\to \tau\tau$ & 5.1 & $H\to \tau\tau$ & 5.1 \\
$\mathcal{C}_{\ell \ell }^{[ijji]}$ & -12. & $m_W$ & 4.5 & $A_b^{\rm FB}$ & 4.5 \\
$\mathcal{C}_{H\ell }^{(3)[33]}$ & -6.8 & $R_\tau$ & 4.5 & $R_\tau$ & 4.5 \\
$\mathcal{C}_{H\ell }^{(1)[33]}$ & -4.6 & $R_\tau$ & 4.4 & $R_\tau$ & 4.4 \\
$\mathcal{C}_{He}^{[33]}$ & -3.8 & $R_\tau$ & 5. & $R_\tau$ & 3.8 \\
$\mathcal{C}_{Hq}^{(1)[33]}$ & -9.7 & $m_W$ & 3.7 & $\Gamma_Z$ & 3.7 \\
$\mathcal{C}_{dH}^{[33]}$ & -4.1 & $H\to bb$ & 3.7 & $H\to bb$ & 3.7 \\
$\mathcal{C}_{Hq}^{(3)[33]}$ & -6.4 & $R_b$ & 3.5 & $R_b$ & 3.5 \\
$\mathcal{C}_{\ell \ell }^{[i33i]}$ & -3.3 & $(g_\tau/g_\mu)_{\tau}$ & 7. & $(g_\tau/g_\mu)_{\tau}$ & 3.3 \\
$\mathcal{C}_{Hu}^{[33]}$ & -3. & $A_b^{\rm FB}$ & 6.6 & $m_W$ & 3. \\
$\mathcal{C}_{\ell q}^{(1)[ii33]}$ & -3.3 & $\sigma_{\rm had}$ & 2.5 & $\sigma_{\rm had}$ & 2.5 \\
$\mathcal{C}_{uB}^{[33]}$ & -4.5 & $m_W$ & 2.4 & $A_b^{\rm FB}$ & 2.4 \\
$\mathcal{C}_{\ell u}^{[ii33]}$ & -2.4 & $\sigma_{\rm had}$ & 3.1 & $\sigma_{\rm had}$ & 2.4 \\
$\mathcal{C}_{\ell q}^{(3)[ii33]}$ & -2.4 & $A_b^{\rm FB}$ & 6.3 & $m_W$ & 2.4 \\
$\mathcal{C}_{Hu}^{[ii]}$ & -4.2 & $\Gamma_Z$ & 2.4 & $R_\tau$ & 2.4 \\
$\mathcal{C}_{uW}^{[33]}$ & -4.5 & $m_W$ & 2.3 & $A_b^{\rm FB}$ & 2.3 \\
$\mathcal{C}_{qq}^{(3)[ii33]}$ & -3.3 & $\Gamma_Z$ & 2.3 & $R_b$ & 2.3 \\
$\mathcal{C}_{\ell q}^{(3)[iijj]}$ & -2.1 & $(g_\tau/g_\mu)_{\pi}$ & 3.5 & $(g_\tau/g_\mu)_{\pi}$ & 2.1 \\
$\mathcal{C}_{\ell equ}^{(1)[3333]}$ & -2.1 & $H\to \tau\tau$ & 3.2 & $H\to \tau\tau$ & 2.1 \\
$\mathcal{C}_{eu}^{[ii33]}$ & -3. & $\sigma_{\rm had}$ & 2.1 & $\sigma_{\rm had}$ & 2.1 \\
$\mathcal{C}_{eq}^{[ii33]}$ & -2.1 & $\sigma_{\rm had}$ & 3.1 & $\sigma_{\rm had}$ & 2.1 \\
$\mathcal{C}_{\ell q}^{(3)[33ii]}$ & -4.4 & $(g_\tau/g_\mu)_{\pi}$ & 2. & $(g_\tau/g_\mu)_{\pi}$ & 2. \\
$\mathcal{C}_{Hq}^{(1)[ii]}$ & -3.1 & $R_c$ & 1.9 & $R_c$ & 1.9 \\
$\mathcal{C}_{quqd}^{(1)[3333]}$ & -1.9 & $H\to bb$ & 1.7 & $H\to bb$ & 1.7 \\
$\mathcal{C}_{Hd}^{[ii]}$ & -1.7 & $R_\tau$ & 2.9 & $\Gamma_Z$ & 1.7 \\
$\mathcal{C}_{qq}^{(1)[3333]}$ & -1.6 & $R_b$ & 3.9 & $m_W$ & 1.6 \\
$\mathcal{C}_{qq}^{(3)[iijj]}$ & -1.5 & $R_\tau$ & 2.3 & $\Gamma_Z$ & 1.5 \\
$\mathcal{C}_{\ell u}^{[3333]}$ & -1.4 & $R_\tau$ & 1.4 & $R_\tau$ & 1.4 \\
 \hline
\end{tabular}
}
 \caption{Bounds from electroweak observables exceeding 1.5~TeV.}
 \label{tab:EW}
 \end{table}

 \newpage 
 
    \begin{table}[t]
     \centering
    \small{
\renewcommand{\arraystretch}{1.25} 
\begin{tabular}{c|c|c|c|c|c}
coeff. & $\Lambda_{-}\text{ [TeV]}$ & Obs. & $\Lambda_{+}\text{ [TeV]}$ & Obs. & $\Lambda\text{ [TeV]}$\\ \hline\hline
$\mathcal{C}_{\ell q}^{(3)[iijj]}$ & -22.5 & $pp\to\mu\nu$ & 26.9 & $pp\to e\nu$ & 22.5 \\
$\mathcal{C}_{\ell q}^{(3)[33ii]}$ & -8.6 & $pp\to\tau\nu$ & 12.4 & $pp\to\tau\nu$ & 8.6 \\
$\mathcal{C}_{eu}^{[iijj]}$ & -9.7 & $pp\to\mu\mu$ & 7.4 & $pp\to ee$ & 7.4 \\
$\mathcal{C}_{eq}^{[iijj]}$ & -8.1 & $pp\to\mu\mu$ & 6. & $pp\to\mu\mu$ & 6. \\
$\mathcal{C}_{\ell q}^{(1)[iijj]}$ & -9. & $pp\to\mu\mu$ & 5.4 & $pp\to\mu\mu$ & 5.4 \\
$\mathcal{C}_{\ell u}^{[iijj]}$ & -7.9 & $pp\to\mu\mu$ & 5.2 & $pp\to\mu\mu$ & 5.2 \\
$\mathcal{C}_{\ell d}^{[iijj]}$ & -4.7 & $pp\to\mu\mu$ & 5.7 & $pp\to\mu\mu$ & 4.7 \\
$\mathcal{C}_{\ell \ell }^{[ijji]}$ & -4.5 & $(e^+e^-\to\mu^+\mu^-)_{\rm FB}$ & 4.4 & $(e^+e^-\to\mu^+\mu^-)_{\rm FB}$ & 4.4 \\
$\mathcal{C}_{\ell \ell }^{[iijj]}$ & -4.5 & $(e^+e^-\to\mu^+\mu^-)_{\rm FB}$ & 4.4 & $(e^+e^-\to\mu^+\mu^-)_{\rm FB}$ & 4.4 \\
$\mathcal{C}_{ed}^{[iijj]}$ & -4.4 & $pp\to\mu\mu$ & 6.5 & $pp\to\mu\mu$ & 4.4 \\
$\mathcal{C}_{ee}^{[iijj]}$ & -4.3 & $(e^+e^-\to\mu^+\mu^-)_{\rm FB}$ & 4.2 & $(e^+e^-\to\mu^+\mu^-)_{\rm FB}$ & 4.2 \\
$\mathcal{C}_{\ell e}^{[iijj]}$ & -5.9 & $(e^+e^-\to\mu^+\mu^-)_{\rm FB}$ & 3.8 & $(e^+e^-\to\mu^+\mu^-)_{\rm FB}$ & 3.8 \\
$\mathcal{C}_{H\ell }^{(3)[ii]}$ & -3.8 & $(e^+e^-\to\mu^+\mu^-)_{\rm FB}$ & 3.9 & $(e^+e^-\to\mu^+\mu^-)_{\rm FB}$ & 3.8 \\
$\mathcal{C}_{eq}^{[33ii]}$ & -5.1 & $pp\to\tau\tau$ & 3.7 & $pp\to\tau\tau$ & 3.7 \\
$\mathcal{C}_{uW}^{[33]}$ & -3.6 & QuarkDipoles & 4. & QuarkDipoles & 3.6 \\
$\mathcal{C}_{\ell q}^{(1)[33ii]}$ & -5.8 & $pp\to\tau\tau$ & 3.4 & $pp\to\tau\tau$ & 3.4 \\
$\mathcal{C}_{\ell \ell }^{[ii33]}$ & -4.6 & $(e^+e^-\to\mu^+\mu^-)_{\rm FB}$ & 3.4 & $(e^+e^-\to\mu^+\mu^-)_{\rm FB}$ & 3.4 \\
$\mathcal{C}_{\ell \ell }^{[i33i]}$ & -4.6 & $(e^+e^-\to\mu^+\mu^-)_{\rm FB}$ & 3.3 & $(e^+e^-\to\mu^+\mu^-)_{\rm FB}$ & 3.3 \\
$\mathcal{C}_{ee}^{[ii33]}$ & -4.4 & $(e^+e^-\to\mu^+\mu^-)_{\rm FB}$ & 3.2 & $(e^+e^-\to\mu^+\mu^-)_{\rm FB}$ & 3.2 \\
$\mathcal{C}_{\ell u}^{[33ii]}$ & -5.1 & $pp\to\tau\tau$ & 3.1 & $pp\to\tau\tau$ & 3.1 \\
$\mathcal{C}_{Hq}^{(3)[ii]}$ & -3.1 & $pp\to e\nu$ & 3.7 & $pp\to\mu\nu$ & 3.1 \\
$\mathcal{C}_{\ell d}^{[33ii]}$ & -3. & $pp\to\tau\tau$ & 3.7 & $pp\to\tau\tau$ & 3. \\
$\mathcal{C}_{H\ell }^{(1)[ii]}$ & -3.3 & $(e^+e^-\to\mu^+\mu^-)_{\rm FB}$ & 3. & $(e^+e^-\to\mu^+\mu^-)_{\rm FB}$ & 3. \\
$\mathcal{C}_{eu}^{[33ii]}$ & -6.8 & $pp\to\tau\tau$ & 2.8 & $pp\to\tau\tau$ & 2.8 \\
$\mathcal{C}_{ed}^{[33ii]}$ & -2.8 & $pp\to\tau\tau$ & 4.2 & $pp\to\tau\tau$ & 2.8 \\
$\mathcal{C}_{uG}^{[33]}$ & -6.7 & QuarkDipoles & 2.7 & QuarkDipoles & 2.7 \\
$\mathcal{C}_{He}^{[ii]}$ & -2.7 & $(e^+e^-\to\mu^+\mu^-)_{\rm FB}$ & 3.2 & $(e^+e^-\to\mu^+\mu^-)_{\rm FB}$ & 2.7 \\
$\mathcal{C}_{qq}^{(3)[i33i]}$ & -2.4 & FourQuarksTop & 2.6 & FourQuarksTop & 2.4 \\
$\mathcal{C}_{qq}^{(3)[ii33]}$ & -2.5 & FourQuarksTop & 2.1 & FourQuarksTop & 2.1 \\
$\mathcal{C}_{\ell e}^{[ii33]}$ & -2. & $(e^+e^-\to\mu^+\mu^-)_{\rm FB}$ & 2.9 & $(e^+e^-\to\mu^+\mu^-)_{\rm FB}$ & 2. \\
$\mathcal{C}_{uB}^{[33]}$ & -1.9 & QuarkDipoles & 2.4 & QuarkDipoles & 1.9 \\
$\mathcal{C}_{\ell e}^{[33ii]}$ & -1.9 & $(e^+e^-\to\mu^+\mu^-)_{\rm FB}$ & 2.9 & $(e^+e^-\to\mu^+\mu^-)_{\rm FB}$ & 1.9 \\
$\mathcal{C}_{eW}^{[33]}$ & -1.9 & $pp\to\tau\nu$ & 1.9 & $pp\to\tau\nu$ & 1.9 \\
$\mathcal{C}_{H\ell }^{(3)[33]}$ & -2.2 & $(e^+e^-\to\mu^+\mu^-)_{\rm FB}$ & 1.7 & $(e^+e^-\to\mu^+\mu^-)_{\rm FB}$ & 1.7 \\
 \hline
\end{tabular}
}
 \caption{Bounds from collider observables exceeding  1.7~TeV.}
 \label{tab:Coll}
 \end{table}

\newpage
\FloatBarrier
 
\bibliographystyle{JHEP}
\bibliography{refs}

\end{document}